\documentclass[article,showpacs,preprintnumbers,amsmath,amssymb]{revtex4-2}

\usepackage{hyperref}

\usepackage{caption}
\usepackage{subcaption}
\usepackage{graphicx}
\begin{document}

%	\title{Charged particles go round and round: ISCO analysis of charged particles in magnetized RN-AdS-global-monopole spacetime}
	\title{Innermost stable circular orbits around a  Reissner-Nordstr\"om-global monopole spacetime in a homogeneous magnetic field}
	%\author{}

	\begin{abstract}
		We investigate the dynamics of charged particles in the spacetime of a global monopole swallowed by a Reissner-Nordstr\"om (RN) black hole in the presence of an external, weak, asymptotically homogeneous magnetic field. We carefully analyze and deduce the conditions to have such a magnetic field around this black hole and show that this is indeed possible in the small but nontrivial charge and monopole term limit. We obtain the general equations of motion and analyze them for special cases of circular orbits, focusing on the innermost stable circular orbit (ISCO) of this configuration. The richness of the parameters and complicated forms of the resulting equations of motion necessitate a numerical approach. Hence, we have presented our results with numerous graphs, which help to understand the evolution of ISCO as a function of the external test magnetic field and the monopole term, depending on the parameters of the black hole, such as its electric charge, as well as the properties of the test particle, such as its specific charge, angular momentum, and energy. We have also analyzed the effective potential that these fields generate and deduced results for the aforementioned values of the external and internal parameters of the spacetime.
	\end{abstract}
	

     \author{Hamza M. Haddad}
	\email{hamza.haddad@tu-dortmund.de}
	\affiliation{Department of Physics, Faculty of Science, Marmara University,  Istanbul, T\" urkiye}
	\date{\today}
    \author{M. Haluk  Se\c cuk}
	\email{haluk.secuk@marmara.edu.tr}
	\affiliation{Department of Physics, Faculty of Science, Marmara University,  Istanbul,  T\" urkiye}
     \author{\"Ozg\"ur Delice}
	\email{ozgur.delice@marmara.edu.tr}
	\affiliation{Department of Physics, Faculty of Science, Marmara University,  Istanbul,  T\" urkiye}
	\date{\today}
	
	\maketitle	
	
	\section{Introduction}

	\label{introduction}

	One of the most important results of General Relativity is its predictions about compact objects called black holes. These black hole solutions lead to intriguing discoveries and concepts, such as singularity theorems,  the concept of the event horizon, black hole thermodynamics, and semiclassical effects such as Hawking radiation, among others \citep{Frolov:1998wf,Misner:1973prb}. These topics and their progress help us increase our knowledge of advanced mathematical techniques and develop new understandings of the fundamental structure of the universe. In the conventional,  classical manner, the relativistic dynamics of test particles  leads to an understanding of the astrophysical behavior of and around compact matter distributions, such as the motion of planets, galaxies, etc., and the behavior of accretion disks around black holes. In the case of accretion disks around black holes, there is an interesting limiting boundary where, in the region closer than this edge, objects rotating around the black hole cannot stay in a stable orbit and should eventually fall into the black hole, whereas objects farther than this region can orbit forever. Therefore, this boundary determines the inner edge of accretion disks in the thin disk limit proposed by Shakura and Sunyaev \citep{Shakura:1972te,Abramowicz:2011xu}. For more general thick disk models, it played an important role in determining bound solutions by limiting the parameter space \citep{Abramowicz:2011xu,Abramowicz:1978}. This limiting boundary is called the innermost stable circular orbit (ISCO). In this paper, we investigate how the ISCO is related to the parameters of a particular black hole spacetime.

The existence and production of magnetic fields near black holes were proposed with the help of several ideas \citep{Frolov:1998wf,Zakharov_2003}. One idea is that magnetic fields can be produced during the motion of charged particles (conducting plasma) in the form of an accretion disk moving around black holes. Another idea is that remnant charges and magnetic fields can be acquired by  black holes during their formation process from the collapsing stars that had magnetic fields. Hence, it might be important to discuss the possible effects of these magnetic fields on the black hole and the particles around it. In this regard, one of our aims is to study the consequences of magnetic fields on the ISCO around  static spherically symmetric black holes carrying an electrical charge with a global monopole.

Global monopoles \citep{Barriola:1989hx} are point topological defects  believed to have been produced during the early universe in the process of symmetry-breaking phase transitions in general unified theories \citep{Kibble:1976sj,Vilenkin:1984ib}. They might form when a global $O(3)$ symmetry is spontaneously broken to $U(1)$. They have some very peculiar gravitational effects, such as the spacetime around them carrying a solid angle deficit, their active gravitational mass vanishes, and their core has tiny repulsive effects \citep{Harari:1990cz}. When they were first studied on  astrophysical scales, they were thought to be one of the \textit{seeds} contributing to the density fluctuations leading to structure formation in the early universe \citep{Bennett:1990xy}. However, when the apparatus observing cosmic microwave background radiation was enhanced and new techniques were added, it was realized that the density fluctuations leading to large-scale structures could not have originated from topological defects alone.  They are, however, still not completely ruled out in this regard. Indeed, they can be secondary sources of such fluctuations \citep{Pogosian:2003mz,Bevis:2006mj}, with  a contribution of up to a few percent. Hence, understanding the physical effects of these global monopoles and their interaction with surrounding fields is still an important topic to investigate. For this reason, we consider the dynamics around a global monopole swallowed by a Reissner-Nordstr\"om (RN) black hole potentially carrying a small test magnetic field. For further properties of global monopoles, see \citep{PhysRevD.44.3152,PhysRevD.66.107701,Bertrand_2003,OzgurDelice_2003,Se_uk_2020,haluk} and references therein.

In this paper, we examine the orbits of charged particles of  Reissner-Nordstr\"om global monopole spacetime in a constant magnetic field, with a focus on ISCO. When a black hole is embedded in an external magnetic field, the motion of charged particles in this spacetime becomes much more complex due to the interaction between the particles, the black hole's charge, and the external magnetic field. Our approach involves deriving the equations of motion for charged particles in this spacetime and then numerically analyzing the ISCO behavior of charged test particles under varying strengths of the external magnetic field and the charge of the global monopole. We then analyze the results to determine the conditions under which stable orbits exist and how these conditions are influenced by the physical parameters of the system. Specifically, we investigate how the ISCO is affected by the strength of the external magnetic field perpendicular to the orbit of rotation and the variation of the charge of the global monopole. We aim to expand the analysis done in \citep{Schroven_2021} for the ISCO of charged particles in RN-spacetime, with the inclusion of global monopole charge and an external magnetic field. Previous studies of charged particle motion around magnetized black hole configurations only considered neutral (Schwarzschild or Kerr) black holes and ignored the possibility of the black hole having a nontrivial electric charge. Hence, one aim of our work is to extend the analyses of previous works to nontrivial, albeit small, charged black holes, as we will discuss in the next section. Here, we will  consider only a weak external magnetic field, the strength of which would not affect the curvature of spacetime.

The motion of neutral or charged test particles around black holes is a topic frequently studied in the literature. Here, we provide an incomplete list of references that an interested reader might find useful. For magnetized black holes \cite{Wald:1974np}, the earlier works \citep{Prasanna:1978vh,Frolov:2010mi,ANAliev_1989,AlievANandOzdemirN} investigate the motion of charged particles around magnetized black holes. See \citep{Baker:2023gdc} for a more complete list of references, including the latest works on this subject. Recently, the ISCO of the RN spacetime without a test magnetic field or monopole \citep{Schroven_2021}, Schwarzschild space-time with a weak charge and magnetic field \citep{Hackstein:2019msh,AlZahrani:2021jxe}, and Kerr in a weak magnetic field without \citep{AlZahrani:2014dfi} or with \citep{AlZahrani:2022fas} a test charge are discussed. Note that the recent works \citep{Hackstein:2019msh,AlZahrani:2021jxe,AlZahrani:2022fas} also aim to consider the ISCO of charged particle motion in charged and magnetized Schwarzschild or Kerr black holes. However, in those works, only the test charge limit is considered, where the charge is a test charge around a neutral black hole and does not enter the spacetime metric. In this work, we take a step further and consider a charged black hole where the electrical charge enters the metric and affects the geodesics of neutral particles and the trajectories of charged particles around it.  The prices we will pay to have such a configuration are the following: the nontrivial charge of the black hole should be small, and the resulting equations become much more complicated than for neutral black holes, where the charge is not present in the metric; obtaining analytical solutions is much more difficult and requires numerical methods. As far as we know, our work is the first to consider such a configuration (ISCO of a nontrivially charged black hole in a test magnetic field) and hence explores new territories regarding the motion of particles around black holes.

The paper is organized as follows: The background global monopole geometry is reviewed, and a magnetic field is presented in Section II.  Equations of motion for the charged test particles will be discussed in Section III. In the next section, we will present the equations leading to ISCO and discuss their various aspects. In Section IV, the ISCO of this spacetime will be analyzed in detail with the help of several figures obtained using a numerical approach to the problem. The last section before the conclusion will be devoted to a study of the effective potential that the charged test particles feel due to the external magnetic field and the gravitational field of the Reissner-Nordstr\"om-global monopole spacetime. We end the paper with a brief discussion.

\section{SPACETIME GEOMETRY}
\subsection{Review of Background Geometry}

	Global monopoles  are formed when a global $O(3)$ symmetry is spontaneously broken to $U(1)$. The simplest Lagrangian representing a global monopole can be written as $\mathcal{L}=-1/2\,\partial_\mu\phi^{a}\,\partial^\mu\phi^a-\lambda /4(\phi^a\phi^a-\eta^2)^2$, where $\phi^a$, $a=1,2,3$, is a triplet of the scalar field, and $\eta$ is its vacuum expectation value. For a global monopole, a solution ansatz $\phi^a=f(r)\,x^a/r$, with $x^a x^a=r^2$, is employed. It was shown by Barriola and Vilenkin \citep{Barriola:1989hx} that outside the core, we can take $f\approx 1$. For a static, spherically symmetric line element ansatz
\begin{equation}\label{line element}
	ds^2=-\frac{\Delta}{r^2}dt^2+\frac{r^2}{\Delta}dr^2+r^2\left(d\theta^2+\sin^2\theta d\phi^2\right),
\end{equation}
this leads to the phenomenological global monopole energy-momentum tensor, with only nonvanishing components having \citep{Barriola:1989hx}
\begin{equation}\label{monopoleemt}
	T^0_0\approx T^r_r\approx -\frac{\eta^2}{r^2}.
\end{equation}

An exact solution of the Einstein field equations representing a global monopole outside the core, with the energy-momentum tensor (\ref{monopoleemt}) (not the full Einstein-scalar system), was also found by Barriola and Vilenkin \citep{Barriola:1989hx} where the solution reads (we use natural units where $G=c=1$ and a  mostly plus metric signature) $\Delta=(1-8\pi \eta^2)r^2-2 M r$, where $M=M_{core}=\lambda^{-1/2}\eta$ is the core mass of the monopole, which was also argued that the core mass can be ignored in astrophysical settings. This solution can also represent a global monopole swallowed by a Schwarzschild black hole if $M$ represents the mass of the black hole. Later, it was discussed in \cite{PhysRevD.44.3152} that an exact solution of the Einstein-Maxwell field equations representing a global monopole swallowed by a Reissner-Nordstr\"om black hole is possible, in the presence of the global monopole energy-momentum tensor (\ref{monopoleemt}). 
Namely, the vector potential  and the  line element in the form (\ref{line element}) for a global monopole swallowed by a static, charged black hole can be written as:

\begin{equation}
	\label{vectorpotential}
	A_E=\frac{Q}{r} dt,
\end{equation}
and
\begin{equation} \label{Delta}
	\Delta(r)=b^2\,r^2-2M r+Q^2,
\end{equation}
$M$ is the mass, and $Q$ is the  electrical charge of the black hole,  $b^2=1-8\pi \eta^2$ is the topological factor of the global monopole presented by Barriola and  Vilenkin \citep{Barriola:1989hx}, and $\eta$ is the monopole charge.   In the case $b=1(\eta=0)$, this line element describes Reissner-Nordstr\"om black hole. The inner horizon ($r_-$) and outer (event) horizon ($r_+$) of the black  hole are given by 
\begin{eqnarray}\label{horizon}
	r_{\pm} = \frac{1}{2b^2}\Big(r_{s} \pm\sqrt{r_{s}^2 - 4b^2Q^2}\Big),
\end{eqnarray}
where $r_{s} = 2M$ is the Schwarzschild radius. This shows that we should have $b^2Q^2\le M^2$ to have a black hole, whereas in the extreme case $b^2Q^2=M^2$, we have a single horizon. For $b^2Q^2 > M^2$, we have a naked singularity. We will not consider this possibility in the following discussion and use the physical limit $b^2 Q^2\le M^2$. 

\subsection{Weak test magnetic field}
	In order to determine a magnetic field that is  sufficiently small to be identified as a test field, we should first mention the limit value beyond which it is no longer small. This limit is well known; see, for example, \citep{ANAliev_1989}. If $B_M=1/M\approx 2.4*10^{19}M_\odot/M\, (\text{G})$, where $M_\odot$ is the mass of the sun,  then the magnetic field strongly influences the line element near the horizon of a black hole. Hence, if $B\ll B_M$,  the magnetic field no longer affects the background geometry and can be treated as a weak test field. In most realistic astrophysical scenarios, this limit is obeyed. For example, even near a magnetar, $B\approx 10^{15} (\text{G})$.  In the presence of such a small magnetic field, the curvature elements and motion of neutral test particles are not affected by this magnetic field. This is the limit we consider in this paper. The behavior of the charged test particles is still affected by even such a small magnetic field. The reason for this is that for charged test particles, such as protons or electrons, the charge per unit mass, denoted $q=q_e/m_e$ for electrons, is a large number, namely $10^{21}$ for electrons and about $10^{17}$ for protons \citep{PhysRevD.96.063015}.  Therefore, the smallness of the test magnetic field can be compensated for by a large value for the charge per unit mass of the charged test particles. In summary, although the weak test magnetic field does not affect the motion of neutral test particles, it can dramatically affect that of charged particles.   

Here we consider the presence of a weak magnetic field, which is homogeneous at spatial infinity along a particular (let's say $z$) direction. We also assume that this magnetic field is  {\emph{weak}} in the sense, as we have discussed in the previous paragraph, that it does not affect the background geometry but can still affect the charged particle trajectories.  Previous studies on magnetic fields for astronomically relevant configurations revealed that \citep{Piotrovich:2010aq} %\citep{piotrovich2010magneticfieldsblackholes},
their relative strength compared to curvature parameters, i.e., $M$ and $Q$, is still very weak \citep{Frolov:2010mi}. Hence, a perturbation is enough to reveal the effects of  magnetic fields on the local behavior of a given compact source, such as a black hole or a galaxy. However, as shown in detail in \citep{Frolov:2010mi}, we cannot neglect the effects of a weak magnetic field on the motion of charged particles, since its effect on the motion of electrons or protons is not small enough to be ignored. The reason is that the dominant term in particle motion satisfies $q B \gg B^2$, which is a massive number for the specific charge $q$ (charge per unit mass) with respect to electrons and protons. Note that there are also exact solutions, for example, the Melvin magnetic universe \citep{PhysRev.139.B225}, and its several extensions, such as black holes embedded in a magnetic universe \citep{10.1063/1.522781,Gibbons_2013}. Applications of these solutions to astronomical phenomena are limited, since a test-field approach can be enough to represent the effects of the magnetic fields.
Thus, motivated by the previous work, we study the effects of an external magnetic field merely as a small test field, which affects the behavior of the particles moving in spacetime but does not affect the motion of neutral particles and does not disturb the curvature of  spacetime.

There is a powerful procedure for generating electromagnetic fields from symmetries of the geometry involving Killing vectors of the spacetime, as presented by Wald \citep{Wald:1974np}. A vector $\xi^\mu$ is a Killing vector if it satisfies the Killing equations $\xi_{\mu;\nu}+\xi_{\nu;\mu}=0$. Here, a semicolon denotes a covariant derivative. It was also shown in \citep{Wald:1974np} that it is possible to obtain the following useful result: $\xi^{\mu;\nu}_{\phantom{aaa};\nu}=R^\mu_{\phantom{m}\lambda}\xi^\lambda$, where $R_{\mu\nu}$ is the Ricci tensor of the spacetime. Hence, in vacuum ($R_{\mu\nu}=0$), the Killing vector $\xi^\mu$ behaves like a vector potential $A^\mu$ satisfying source-free Maxwell equations $F^{\mu\nu}_{\phantom{mu};\nu}=0$ where $F_{\mu\nu}=A_{\nu,\mu}-A_{\mu,\nu}$. Since the background geometry we consider is not a vacuum spacetime, ($R_{\mu\nu}\neq 0$), we cannot apply this method to our problem in general. In other words, the external electromagnetic field does not satisfy the background source-free Maxwell equations. However, if the right-hand side can be neglected, we can still use this method to generate the desired external magnetic field. This is discussed in \citep{ANAliev_1989}, for the Kerr-Newman background and suggests that if $Q^2\ll M$, then the background Kerr result is still applicable. Here we have two distinct sources of gravity, $Q$ and $\eta$. Hence, if $Q^2\ll M$ and $\kappa \eta^2 \ll M$, we can take $R_{\mu\nu}\xi^\nu\simeq 0$ and use the vector potential for the background Schwarzschild spacetime as the vector potential. Namely, the Killing vector $\xi^\mu$ can still be considered  a vector potential as long as $Q^2\ll M$ and $\kappa \eta^2 \ll M$.  Note that in the literature, some works that claim to study charged particle motion for a weakly magnetized RN black hole or its generalizations without checking whether the background Maxwell equations are satisfied \citep{Mandal:2023bgw,Majeed:2014kka}. Unfortunately, as we have shown in the above discussion, they are not.

Therefore, we consider an additional {\emph{constant}} vector potential representing  an external test magnetic field, as \citep{ANAliev_1989,Wald:1974np,Baker:2023gdc},
\begin{equation}\label{testmagnetic}
	A^\mu=\frac{B}{2}\,\xi^\mu_{(\phi)},  %\partial_\phi.	A_M=\frac{B}{2}r^2\sin^2\theta\, d\phi, 
\end{equation}
where we suppose that the strength of the magnetic field is small compared to the mass and charge of the black hole  $B\ll B_M$, $B\ll Q$, and also the global monopole term, $B\ll \kappa \eta^2$. This ensures that the external magnetic field does not affect the background geometry.   Since the magnetic field is merely a test field,  it should satisfy the Maxwell equations $dF=0,$ and $d*F=0$, with $F=dA$, of the background geometry (\ref{line element}), at least under the approximations used.   It is trivial to show that indeed $dF=0$ is satisfied, whereas $d*F=0$ is satisfied, as long as our approximation $B\ll B_M$, $Q^2\ll M^2$, and $\kappa \eta^2 \ll M^2$ is valid. Note that, for the Schwarzschild case, this test field satisfies the background Maxwell equations exactly.   

In order to see explicitly that the test field (\ref{testmagnetic}) is indeed a general test field satisfying the Maxwell equations for the background  geometry under the assumptions employed, let us consider the following vector potential one-form
\begin{equation}
	A=h(r)\,dt+g(r)\,sin^2\theta\, d\phi,
\end{equation}
which respects the symmetries of the spacetime, since $\xi^\mu_{(t)}=\partial/\partial t$ and $\xi^\mu_{(\phi)}=\partial/\partial \phi $ are Killing vectors of the spacetime. This vector potential leads to the Faraday one-form via $F=dA$, as
\begin{equation}
	F=-h' dt\wedge  dr +g' \sin^2 \theta dr\wedge d\phi+ 2 g \sin\theta \cos \theta\, d\theta \wedge d\phi.
\end{equation}	

The Maxwell equation $dF=0$ is trivially satisfied, whereas $d*F=0$ (here $*$ is the Hodge dual operator) yields
\begin{eqnarray}\label{Maxwel1}
	h''-2\frac{h'}{r}=0,\\	
	\left(\frac{g'\Delta}{r^2}\right)'-2\frac{g}{r^2}=0.\label{Maxwel2}
\end{eqnarray}
Here, prime means derivative with respect to the radial coordinate. The first equation (\ref{Maxwel1}) has the solution $h=k/r$ with $k$ being a constant, and we can take $k=Q$, the total charge of the black hole, including any small test charge present in the black hole as well. Unfortunately, the second equation (\ref{Maxwel2}) does not have a simple solution.  Even for vanishing $Q$, its solution involves hypergeometric and Meijer $G$-functions, which will make an analytic investigation of particle trajectory equations non-tractable. However, in the small $Q$ and small $\kappa \eta^2$ limit, we can obtain a simple approximate solution as follows. The second equation (\ref{Maxwel2}) involves the product of $\Delta$ and $g'$.
Explicitly, if $Q^2g'\ll g'$ and $\kappa \eta^2 g'\ll g'$, the term inside the bracket becomes
\begin{equation}
	\left(Q^2-2M r+r^2b^2\right) g' \approx \left(-2M r+r^2 \right) g'.
\end{equation}

Under these limits,  from equations (\ref{Maxwel1},\ref{Maxwel2}) we have obtained the following solutions
\begin{eqnarray}
	&&h(r)=\frac{Q}{r},\\
	&&g(r)\approx \frac{B}{2} r^2+ \mathcal{O}(Q^2 B)+\mathcal{O}(\kappa \eta^2 B).
\end{eqnarray}  
From these, we see that, if $Q^2B\ll B$ and $\kappa \eta^2 B\ll B$, we can take the same test magnetic field as in the Schwarzschild case, given in equation (\ref{testmagnetic}).
As a result, the total vector potential that we will use for the rest of the paper reads:
\begin{equation}
	A=\frac{Q}{r} dt+\frac{B}{2}r^2\sin^2\theta\, d\phi,
\end{equation}
where $B^2\ll B_M^2$, $Q^2B\ll B$, and $\kappa \eta^2 B \ll B$. In order to fulfill these conditions,  we will take $Q^2\le 0.1 M$, and $\kappa \eta^2 \le 0.1M$, whenever the magnetic field is present in the rest of the paper.  We choose $Q,\,B>0$, without loss of generality, in the rest of the paper.

\section{EQUATIONS OF MOTION}

	We use the Hamilton-Jacobi (HJ) approach to derive the equations of motion of a test particle, whose  specific electrical charge is denoted by $q$, with the following equation:
\begin{eqnarray}
	\frac{	\partial S }{\partial \lambda} + \frac{1}{2} g^{\mu\nu}\left( \frac{\partial S}{\partial x^\mu}+q A_\mu\right) \left( \frac{\partial S}{\partial x^\nu}+q A_\nu\right) = 0.
\end{eqnarray}
By respecting the symmetries of the spherically symmetric, static spacetime, we may exploit the cyclic nature of the HJ equation in the coordinates $t$ and $\phi$ to write an action $S$ that reads:
\begin{equation}
	S=\frac{1}{2}\delta\, \lambda -E \, t+L\, \phi +S_1(r,\theta),
\end{equation}
where $\lambda$ is an affine parameter, $E$ is the specific energy, $L$ is the specific angular momentum of the particle \citep{Misner:1973prb} and $\delta$ characterizes the trajectory of the  particle, given by
\begin{eqnarray}\label{delta}
	\delta = -\frac{ds^2}{d\lambda^2}.
\end{eqnarray} 
Here, $\delta=0,\,1$ respectively correspond to massless and timelike particles. The Hamilton-Jacobi equation reads
\begin{eqnarray}\label{action1}
	\begin{aligned}
		\delta-\frac{r^2}{\Delta}\left( -E+\frac{q Q}{r}\right)^2+\frac{\Delta}{r^2}
		\left(\frac{\partial S_1}{\partial r} \right)^2  +  \frac{1}{r^2} \left(\frac{\partial S_1}{\partial \theta} \right)^2  + \frac{1}{r^2\sin^2\theta}\left(L+\frac{1}{2}q B r^2 \sin^2\theta \right)^2=0.
	\end{aligned}
\end{eqnarray}
From this expression, we  find the first integrals of motion as follows:
\begin{eqnarray}
	&&	\dot{t}=\pm\frac{r^2}{\Delta}\left(E-\frac{q\,Q}{r} \right),\label{tdot}\\
	&&	\dot{\phi}=\pm\frac{\left(L+q B\,r^2\sin^2\theta /2 \right)}{r^2\sin^2\theta}. \label{phidot}
\end{eqnarray}
For a charged particle coupled with a test magnetic field, the HJ equation \eqref{action1} is not separable. Therefore, we  restrict ourselves to equatorial trajectories by choosing $\theta=\pi/2$ and we find:
\begin{eqnarray}
	\left(\frac{d S_r}{dr} \right)^2= \frac{r^4}{\Delta^2}\left( E-\frac{q Q}{r}\right)^2 - \frac{r^4}{\Delta^2} \left[ \delta r^2+\left( L+\frac{q B r^2}{2}\right )^2\right].
\end{eqnarray}
Using the definition \eqref{delta}, with the first integrals given by \eqref{tdot}, and \eqref{phidot}, the radial equation of motion of a charged particle with charge $q$ in the test magnetic field $B$ in Reissner-Nordstr\"om-monopole space-time can be written in the form:
\begin{eqnarray}\label{r dot}
	\dot{r}^2 = \left(E-\frac{qQ}{r}\right)^2 - \frac{\Delta}{r^4}\left[\delta r^2+(L+qBr^2 /2)^2\right],
\end{eqnarray}
where overdot represents a derivative with respect to the affine parameter, $\lambda$.

If one wishes, it is also possible to get an expression for the radial orbit in terms of the orbit angle by dividing $\dot{r}^2$ by $\dot{\phi}^2$, obtaining 
\begin{eqnarray}\label{dr dphi}
	%	\begin{aligned}
		\left(\frac{dr}{d\phi}\right)^2 =\frac{r^4}{\left( L+\frac{q B r^2}{2}\right )^2}\left\{\left(E-\frac{qQ}{r}\right)^2   -  \frac{\Delta}{r^4}\left[\delta r^2+\left( L+\frac{q B r^2}{2}\right )^2\right]\right\} \equiv U(r).
		%	\end{aligned}
\end{eqnarray}
The equation given in (\ref{r dot}) may also be used to calculate the effective potential of the orbit by the relation:
\begin{eqnarray}\label{r dot pot}
	\dot{r}^2 = (E - V_{+})(E - V_{-}),
\end{eqnarray}
where the expanded form of this equation gives the following relations:
\begin{eqnarray}
	&&V_{+}+V_{-} = \frac{2qQ}{r},\\
	&&V_{+}V_{-} = \left(\frac{qQ} {r} \right)^2 - \frac{\Delta}{r^4}\left[\delta r^2 +\left( L+\frac{q B r^2}{2}\right )^2\right].
\end{eqnarray}
Therefore, the effective potential is found as
\begin{eqnarray}\label{effpot}
	V_{\pm} = \frac{qQ}{r}\pm \sqrt{\frac{\Delta}{r^4}\left[\delta r^2 +\left( L+\frac{q B r^2}{2}\right )^2\right]},
\end{eqnarray}
whose fully expanded form is 
\begin{eqnarray}\label{expanded potentil}
	\begin{aligned}
		V_{\pm} = & \frac{qQ}{r} \pm 
		\Bigg(\frac{1}{4} b^2 B^2 q^2 r^2+b^2 B L q+b^2 \delta+\frac{b^2 L^2}{r^2} \\ &- \frac{1}{2} B^2 M q^2 r  +\frac{1}{4} B^2 q^2 Q^2  -\frac{2 B L M q}{r} 	+\frac{B L q Q^2}{r^2} \\ &
		-\frac{2 \delta M}{r}+\frac{\delta Q^2}{r^2}-\frac{2 L^2 M}{r^3}+\frac{L^2 Q^2}{r^4} \Bigg)^{1/2}.    
	\end{aligned}
\end{eqnarray}

The expression given above for the effective potential can be used to analyze the stability of the orbits and to classify them,  providing a general idea of their behavior. Note that in the special limit where  both the external  magnetic field  and the global monopole parameter vanish ($\eta, B=0$), the presented effective potential reduces to a well-known expression \citep{Schroven_2021,article},
\begin{eqnarray}
	V_{\pm} = \frac{qQ}{r}\pm \sqrt{ \delta+\frac{L^2}{r^2}-\frac{2 \delta M}{r}+\frac{\delta Q^2}{r^2}-\frac{2 L^2 M}{r^3}+\frac{L^2 Q^2}{r^4}}, \nonumber
\end{eqnarray}
corresponding to the effective potential  obtained for the  Reissner-Nordstr\"om spacetime.

\section{ISCO OF CHARGED PARTICLES}

	We will obtain a set of three equations corresponding to the roots and the first and second derivatives of the effective potential, since ISCO is present at the location of an inflection point.
It is possible to use the effective potential to perform this analysis, but, due to the radical expression, it would be very computationally taxing. Thus, an alternative approach is due. Such an approach might involve analyzing the radial orbit equation, specifically the one given in (\ref{dr dphi}), by setting
\begin{eqnarray}\label{dr 0}
	\left(\frac{dr}{d\phi}\right)^2= 0,\\
	\label{dr 1}   \frac{d \,U(r)}{dr}=0,\\
	\label{dr 2}\frac{d^2\, U(r)}{dr^2} =0, 
\end{eqnarray}
where $U(r)$ is defined in (\ref{dr dphi}). A simultaneous solution of these equations in the variable $r$  will give the ISCO of the particle. For a similar treatment of RN spacetime without a test magnetic field or a global monopole, see \citep{Schroven_2021}.
The explicit forms of the equations (\ref{dr 0}, \ref{dr 1}, \ref{dr 2})  are given below. It is clear that the first equation involves a sixth-order polynomial, given by 
\begin{eqnarray}
	\begin{aligned}\label{R = 0}
		&-r^4 \Big(4 b^2 B L q+4 b^2 \delta+B^2 q^2 Q^2-4E^2\Big)-b^2 B^2 q^2 r^6 \\ & -4 r^2 \Big(b^2 L^2+B L q Q^2+\delta Q^2-q^2 Q^2\Big) +2 B^2 M q^2 r^5\\&+8 r^3 \Big[q \Big(B M L -E Q\Big)+\delta M\Big]+8 L^2 M r-4 L^2 Q^2 =0,
	\end{aligned}
\end{eqnarray}
whereas the second one is a seventh-order polynomial  
\begin{eqnarray}
	\begin{aligned} \label{Rderivative}
		&2 L r \biggl\{q^2 \Big[3 B^2 r^3 \Big(b^2 r-M\Big)-4 Q^2\Big]+4 \delta \Big[r \Big(2 b^2 r-3 M\Big)+Q^2\Big]-8 E^2 r^2+12 E q Q r\biggl\}\\ &+B q r^3 \biggl\{q \Big[B^2 q r^3 \Big(b^2 r-M\Big)-4 E Q r+4 q Q^2\Big]-4 \delta \Big(Q^2-M r\Big)\biggl\}+12 B L^2 q r^2 \Big(b^2 r-M\Big)\\ &-8 L^3 \Big(M-b^2 r\Big)=0,
	\end{aligned}
\end{eqnarray}
and the last one has an eighth-order polynomial 
\begin{eqnarray}
	\begin{aligned}\label{Rsecond}
		&b^2 \Big[B^4 q^4 r^8+8 L \Big(B^3 q^3 r^6-6 B \delta q r^4\Big)+24 L^2 \Big(B^2 q^2 r^4+4 \delta r^2\Big)+32 B L^3 q r^2+16 L^4\Big]\\&-4 \biggl\{\delta \Big[B^2 q^2 r^4 \Big(2 M r-3 Q^2\Big)+16 B L q r^2 \Big(Q^2-2 M r\Big)-4 L^2 \Big(Q^2-6 M r\Big)\Big]\\&+B^2 q^3 Q r^4 \Big(3 q Q-2 E r\Big)-4 B L q r^2 \Big(3 E^2 r^2-8 E q Q r+4 q^2 Q^2\Big)\\&+4 L^2 \Big(6 E^2 r^2-6 E q Q r+q^2 Q^2\Big)\biggl\}=0.
	\end{aligned}
\end{eqnarray} 

What we have obtained  is  a set of three polynomial equations of the form:
\begin{equation}\label{polynomialeq}
	\sum a_{i}r^i=0,
\end{equation}
which can,  in principle, be solved for three unknowns. Since the equations at hand are polynomials of the form (\ref{polynomialeq}), we may analyze the polynomial coefficients $a_{i}$ of  equations (\ref{dr 0}, \ref{dr 1}, \ref{dr 2}). For equation (\ref{dr 0}), we have the following coefficients:
\begin{eqnarray*}
	&&a_{6} = -b^2 (qB)^2 /4,\\
	&&a_{5} = 2M(qB)^2 /4,\\
	&&a_{4} = E^2 -b^2 \delta -b^2 LqB - Q^2 (qB)^2 /4,\\
	&&a_{3} = -2 EqQ+2M\delta +2MLqB,\\
	&&a_{2} = (qQ)^2 -Q^2 \delta -b^2 L^2 - Q^2 LqB,\\
	&&a_{1} = 2ML^2,\\
	&&a_{0} = -Q^2 L^2.
\end{eqnarray*}

A polynomial in the form of $\sum a_{i}r^i$ can have as many roots as the highest power of $r$. We can then employ Descartes's rule of signs: `` An equation can have as many true (positive) roots as it has changes of sign, from $+$ to $-$ or from $-$ to $+$." Hence,  the number of positive roots a polynomial has is equal to the number of sign changes of consecutive, non-zero coefficients $a_{i}$, or less than it by an even number \citep{doi:10.1080/00029890.1999.12005131,ruleOfSings}.
The polynomial given in \eqref{R = 0} can have a maximum of $6$ positive roots. Since $a_{0}$ can only be negative, $a_{1}$ can only be positive, $a_{5}$ can only be positive, and $a_{6}$ can only be negative, it must have at least $3$ sign changes of consecutive coefficients. Therefore, for $B\neq 0$ and $q\neq 0$, we should have either three or one positive roots. To have the maximum number of positive roots, the following restrictions must be satisfied:
\begin{eqnarray*}
	&&a_{4} = E^2 -b^2 \delta -b^2 LqB - Q^2 (qB)^2 /4<0,\\
	&&a_{3} = -2 EqQ+2\delta +2LqB>0,\\
	&&a_{2} = (qQ)^2 -Q^2 \delta -b^2 L^2 - Q^2 LqB<0.
\end{eqnarray*}
Here, for simplicity, the mass of the black hole is set to $M=1$. In our calculations, we will be interested only in the smallest real positive root of the polynomial, since that would be the ISCO for a massive charged particle. Note that for a photon, $q=0$ and $\delta = 0$; thus, the non-zero coefficients for null geodesics are: 
\begin{eqnarray}
	&&a_{4} = E^2,\\
	&&a_{2} = -b^2 L^2 ,\\
	&&a_{1} = 2ML^2,\\
	&&a_{0} = -Q^2 L^2,
\end{eqnarray}
and since $E^2$ cannot be negative, this leaves only two consecutive sign changes. Therefore, for photons, two or zero positive roots can be found. A more convenient approach to obtaining a solution for the photon sphere would be using the effective potential (\ref{effpot}), which simplifies considerably for null-geodesics since we have $\delta =q = 0$.
To obtain the radius of the photon sphere, one might set the first derivative of the effective potential (\ref{effpot}) equal to zero and solve for $r$ at $\delta = 0$ and $q = 0$. The result is:
\begin{eqnarray}\label{photon sphere}
	r_{ps} = \frac{3 M\pm\sqrt{9 M^2-8 b^2 Q^2}}{2 b^2}.
\end{eqnarray}
We observe that the radius of the photon sphere is not affected by the external magnetic field, as expected, since only charged particles will be affected by it. When there is no monopole, which means $b=1$, the photon sphere simplifies to the corresponding expression of the photon sphere in  RN-spacetime, given in \citep{PhotonSpherePhysRevD.104.124016}. The topic of photon orbits will not be discussed further in this paper due to their trivial nature.

Note that the equation (\ref{R = 0}) is a quadratic equation for $E$ and $L$,  where the equation (\ref{Rderivative}) is quadratic for $E$. To find the ISCO radius, we can try to solve (\ref{R = 0}), which has the form  $\alpha\, L^2+\beta\, L+\gamma =0$, for $L$ and insert this result into the other equations. But, this makes the equation \eqref{Rderivative} unmanageable, because now, the resulting equation involves complicated expressions having square roots of the other unknown variables. The same approach also holds for equations \eqref{Rderivative} and \eqref{Rsecond}. Hence, an analytical expression for the present setup is unattainable for the ISCO radius. Due to this difficulty, we need a numerical approach to the problem.
\begin{figure*}[ht]
	\centering
	\begin{subfigure}{.5\textwidth}
		\centering
		\includegraphics[width=.9\linewidth]{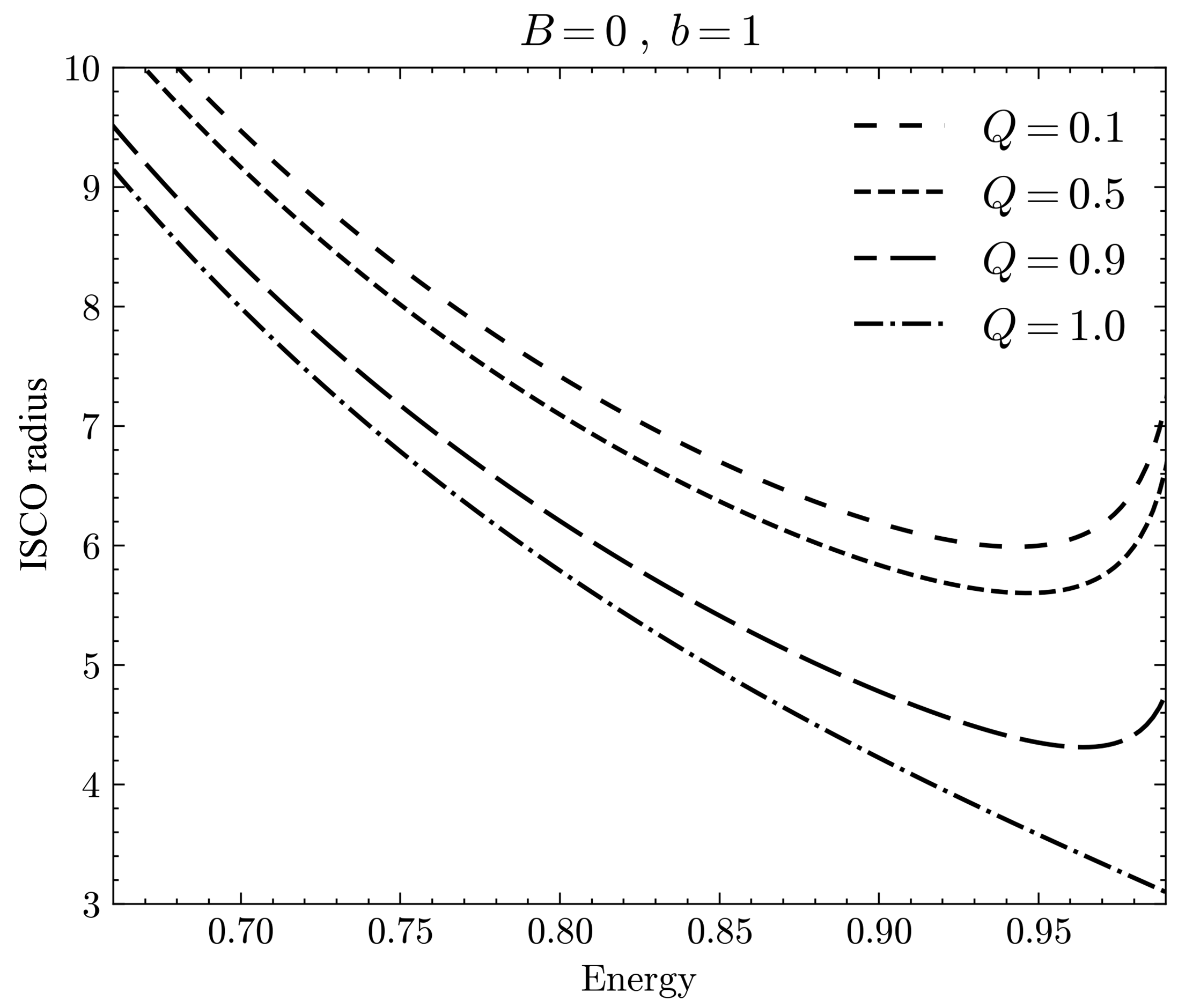}
		\caption{$\eta=0$}
		\label{fig1:sub1}
	\end{subfigure}%
	\begin{subfigure}{.5\textwidth}
		\centering
		\includegraphics[width=.9\linewidth]{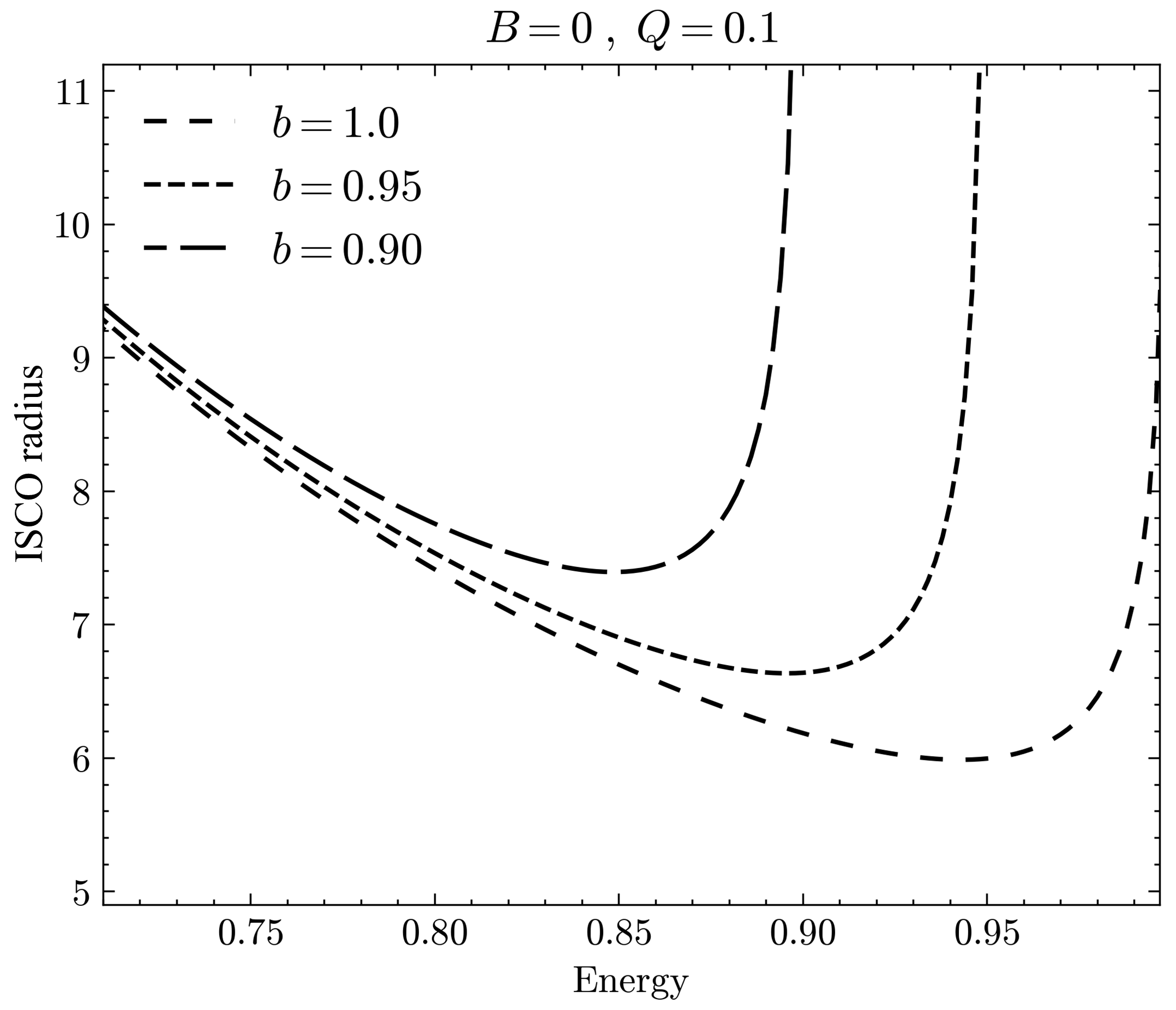}
		\caption{$\eta\neq0$}
		\label{fig1:sub2}
	\end{subfigure}
	\caption{
		The Fig.s (\ref{fig1:sub1}, \ref{fig1:sub2}) summarize the change of ISCO with the energy of the test particle for Reissner-Nordstr\"om black hole with or without a global monopole when there is no external magnetic field. Fig. (\ref{fig1:sub1}) shows the change in ISCO for increasing black hole electric charge without the global monopole, which agrees with the results presented in \citep{Schroven_2021}. In the right, Fig. (\ref{fig1:sub2}), we fixed the black hole charge, $Q=0.1$, to observe the effects of the global monopole on the ISCO curves. We see that by increasing the global monopole parameter, $\eta$, or, equivalently, decreasing $b$, the general behavior does not change. However, the ISCO curves are shifted, so the energy required for a particle to follow a circular orbit increases with increasing  monopole parameter.
	} 
	\label{fig:test1}
\end{figure*}

\section{ANALYSIS OF ISCO}

	The numerical analysis procedure we have considered in this paper is as follows: First, we assign the external parameters ($B$, $Q$, and $b$) to fixed values. Later, a value for the independent parameter ($E$) is specified, and then the system of equations is simultaneously solved for $r$, $L$, and $q$. This process returns a set of solutions for the system of equations (\ref{R = 0},\ref{Rderivative},\ref{Rsecond}). Since we are only interested in the ISCO, the smallest real positive value of $r$ is the solution of interest. This process iterated over the domain of $E \in (0, 1)$ with a step size of $\Delta E\approx 0.001$. Subsequently, this process was repeated for various configurations of the external parameters to explore their effects by plotting the resulting dataset for further analysis.
No solutions were obtained at particular random values of $E$, leading to minor discontinuities in the data. These gaps are attributed to numerical errors and, as a result, were neglected. However, the overall trend remained consistent and unaffected. We will discuss the results of the analysis below in detail. 
\begin{figure*}[ht] 
	\centering
	\begin{subfigure}{.5\textwidth}
		\centering
		\includegraphics[width=.9\linewidth]{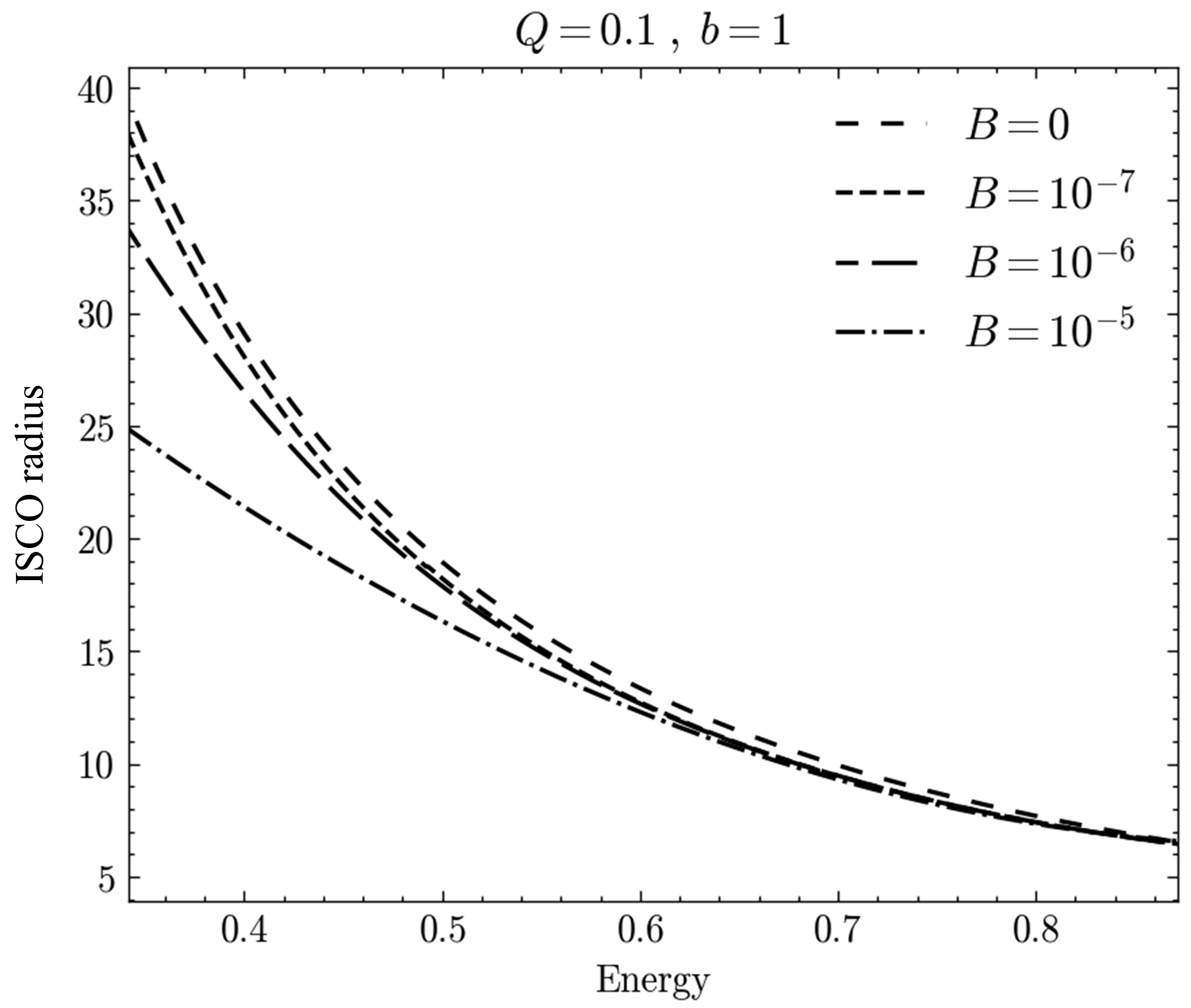}
		\caption{The effect of the magnetic field on ISCO}
		\label{fig2:sub1}
	\end{subfigure}%
	\begin{subfigure}{.5\textwidth}
		\centering
		\includegraphics[width=.9\linewidth]{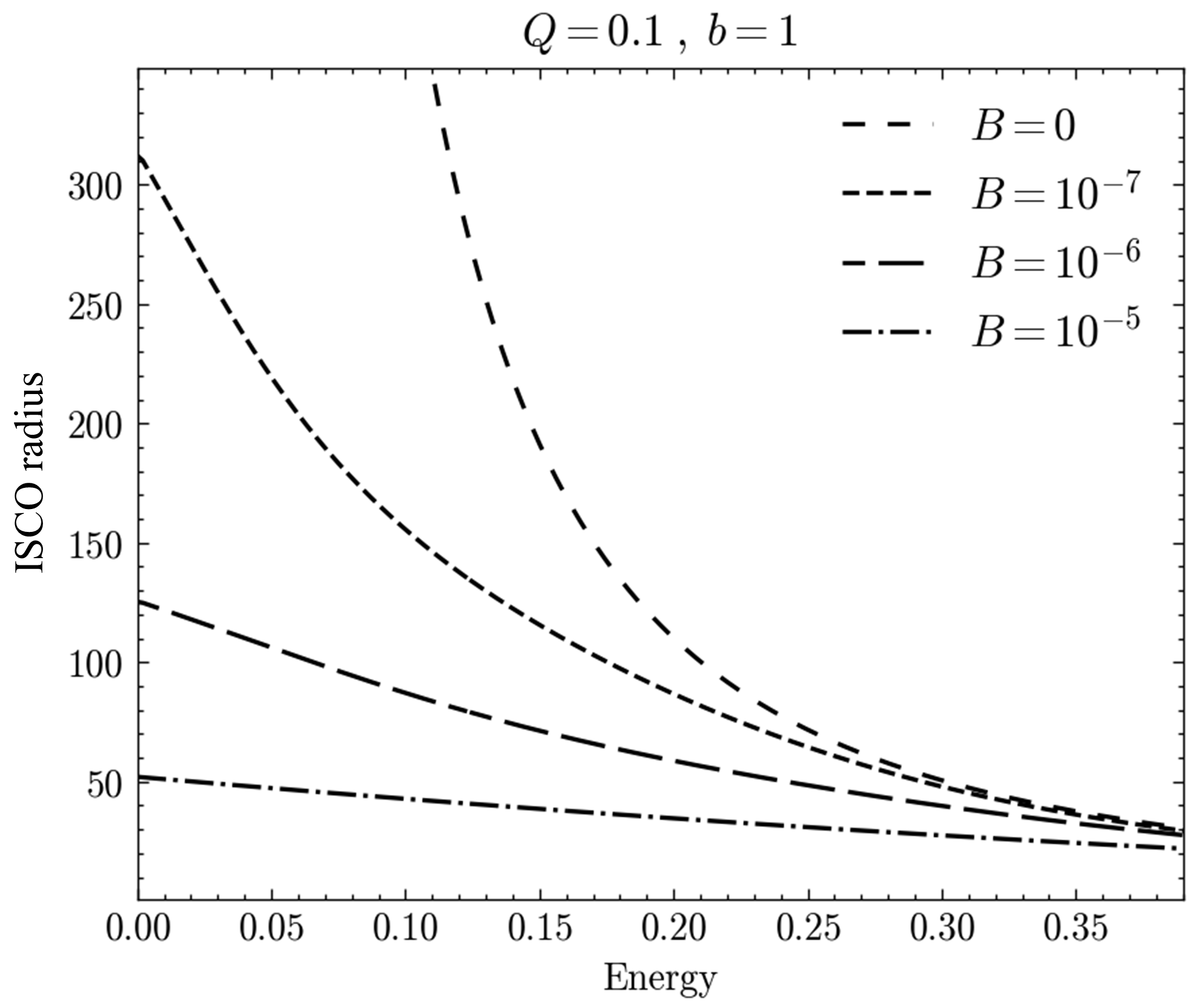}
		\caption{The effect of the magnetic field on ISCO at lower energy values}
		\label{fig2:sub2}
	\end{subfigure}
	\caption{The dependence of ISCO on the energy of the test particle in the presence of a weak magnetic field is plotted for various values of the external magnetic field. When the magnetic field increases, the ISCO radii decrease for a given particle energy. Fig. (\ref{fig2:sub2}) shows that the presence of a magnetic field changes the low-energy limit of the ISCO. Compared to a zero magnetic field, where the ISCO radii approach infinity as the energy approaches zero, the ISCO radii remain finite in this limit, implying a stabilizing effect.}
	\label{fig:2}
\end{figure*}

We have $\delta = 1$ for massive particles and will restrict ourselves to positive energy values. Since we have three equations, we may solve for three unknowns, which we will choose to be $r$, $q$, and $L$. We will only consider real, positive values of $r$ in the solutions. We present our findings using this approach in a series of graphs.

Fig. (\ref{fig1:sub1}) shows the behavior of the innermost stable circular orbit for a charged particle around a Reissner-Nordstr\"om black hole, in the case where there is no external magnetic field, $B = 0$, and a monopole charge $\eta$ is set to zero,  which means $b = 1$. The left figure of this graph is a cross-check for our method, which will be applied to new configurations since we observe the same behavior detailed in \citep{Schroven_2021} for the same case. In this scenario, we see that as the energy of the test particle decreases and approaches zero, the ISCO gradually shifts outward, ultimately going to infinity. The more extreme change occurs as $E\to 1$, where the ISCO reaches its minimum point and then rapidly increases to infinity. An interesting pattern emerges when we generalize this solution to include the monopole parameter with a non-zero monopole charge, as shown in Fig.  (\ref{fig1:sub2}). Namely, the monopole charge effectively shifts the graph to the left, and similarly, the energy at which the ISCO goes to infinity  shifts accordingly. For example, without a monopole in the core of the black hole,  $b = 1$, the ISCO goes to infinity at $E = 1$. As we decrease $b$ to $0.95$, this behavior occurs at a lower energy, specifically at $E = 0.95$. Similarly, when $b = 0.90$, the ISCO goes to infinity at $E = 0.90$.

This behavior is notably independent of the electric charge of the black hole, $Q$, and the external magnetic field, $B$. Regardless of these parameters, the shift in the ISCO due to changes in the value of the monopole parameter remains consistent. Therefore, one can interpret a change in the monopole charge as functionally equivalent to a change in the energy of the test particle. Furthermore, the ISCO curves shift upward when the global monopole charge increases. As a result, we can see that the existence of a monopole has a repulsive effect on a particle's orbit. The higher the monopole charge,  the more a particle is pushed outward. This trend fits with the analysis done in \citep{haluk}, where an increased monopole charge lowered the deflection angle of incoming particles. The decrease in the deflection angle may be related to the repulsive nature of the global monopole\citep{Harari:1990cz}. This behavior is reiterated in Eq. (\ref{photon sphere}), such that when $b$ decreases, the radius of the photon sphere increases.

Having deduced how the ISCO is affected by the existence of a monopole for RN black holes, let us now consider more general cases by including the external test magnetic field. We do this step by step, first discussing the effect of a magnetic field on the charged particles in the RN black hole without a monopole. The behavior of the ISCO is summarized in the Fig.s (\ref{fig1:sub1},\ref{fig1:sub2}) in the presence of a weak external magnetic field as a function of the particle's energy. In Fig. (\ref{fig2:sub1}),  we observe that as the strength of the external magnetic field increases, the ISCO decreases for lower energy values. This indicates that a stronger magnetic field pulls the ISCO closer to the black hole, allowing orbits of smaller radii. This result is in accordance with neutral black holes \citep{ANAliev_1989,AlievANandOzdemirN,ZNAJEK1976,Frolov:2011ea}. However, when the particle energy approaches $E \to 1$, all the different ISCO curves converge to the same point, which is the minimum ISCO value, before going to infinity at $E=1$, showing that the influence of the magnetic field diminishes at higher energy values.
\begin{figure*}[ht]
	\centering
	\begin{subfigure}{.5\textwidth}
		\centering
		\includegraphics[width=.9\linewidth]{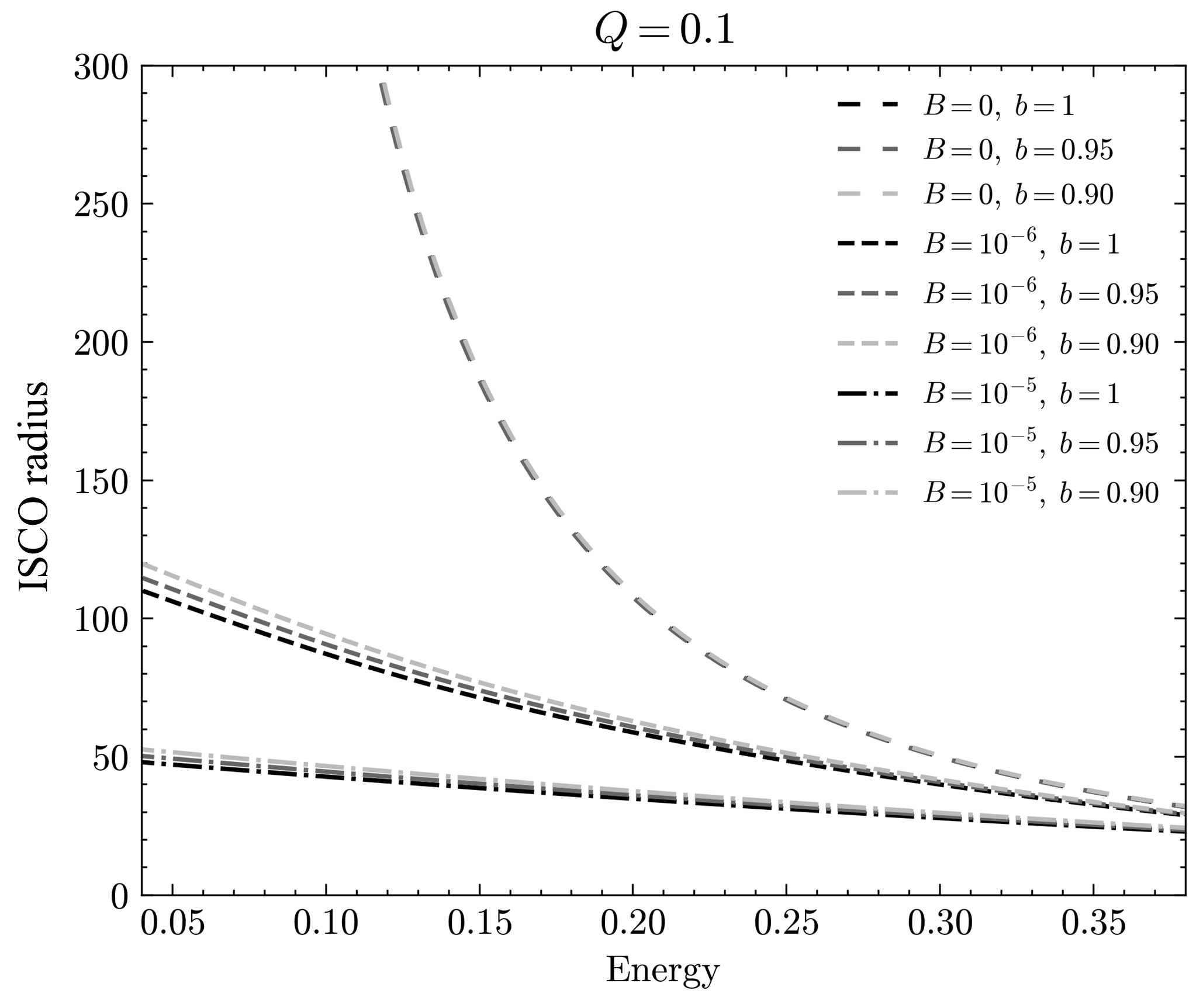}
		\caption{Combined effect of $B$ and $\eta$ on the ISCO}
		\label{B and b:sub1}
	\end{subfigure}%
	\begin{subfigure}{.5\textwidth}
		\centering
		\includegraphics[width=.9\linewidth]{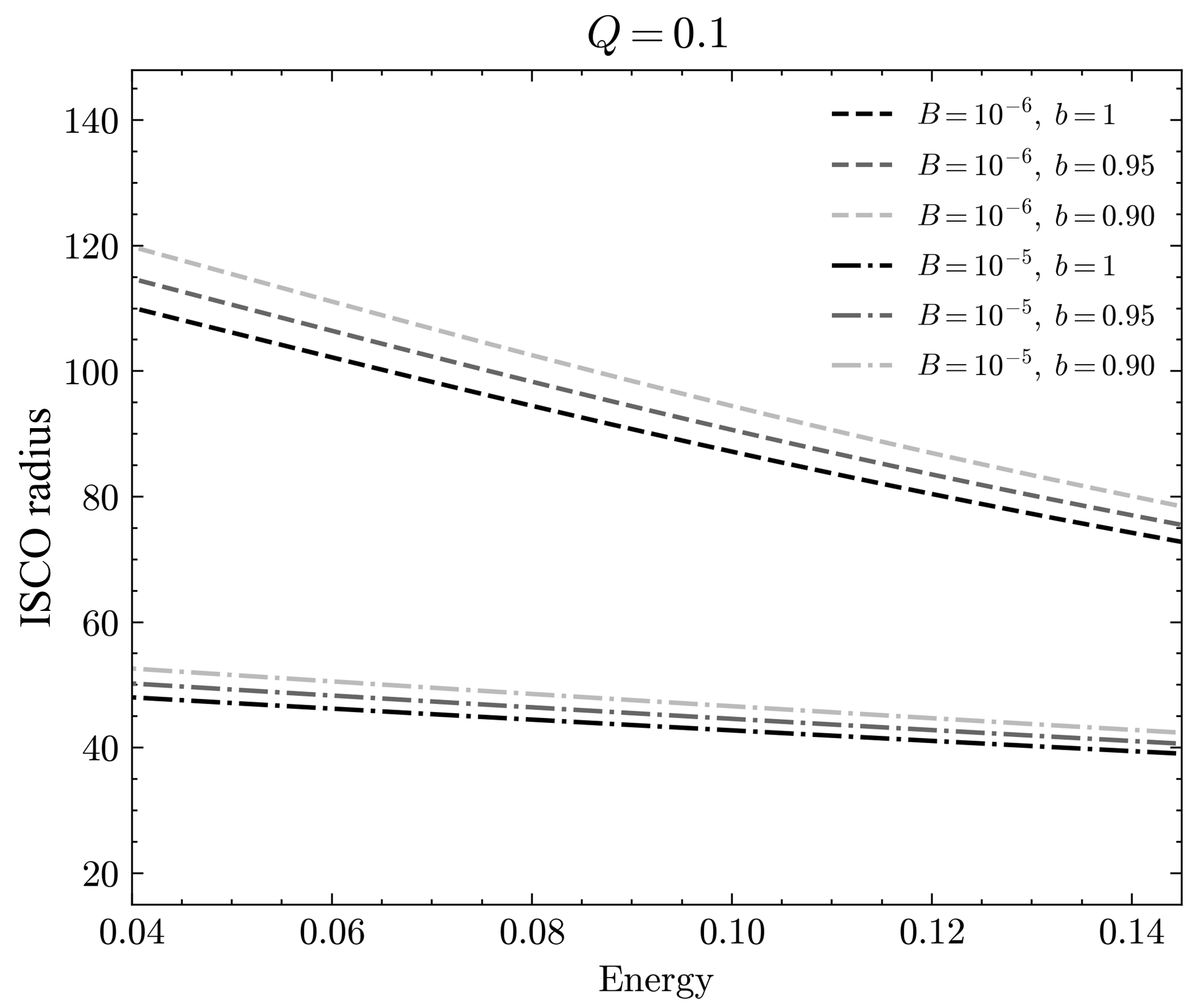}
		\caption{A close inspection of low energy values of (\ref{B and b:sub1})}
		\label{B and b:sub2}
	\end{subfigure}
	\caption{Effect of a  monopole and a magnetic field on the ISCO. The magnetic field forces the ISCO radii closer to the black hole, Fig. (\ref{B and b:sub1}), whereas the global monopole forces them further away for a given energy of a test particle, Fig. (\ref{B and b:sub2}).}
	\label{fig:3}
\end{figure*}

In Fig. (\ref{fig2:sub2}), the behavior of the ISCO as the energy approaches zero is highlighted. Unlike the case without a magnetic field, where the ISCO goes to infinity at very low energies, a magnetic field changes this behavior. For  $B \neq 0$, the ISCO does not diverge to infinity as $E \to 0$. Instead, the magnetic field imposes a limiting effect, causing the ISCO to reach a maximum value when $E = 0$. This indicates that even a weak external magnetic field can sustain stable orbits at very low energy values, preventing the ISCO from receding outward indefinitely. This behavior shows a stabilizing influence of the magnetic field. From  Figures (\ref{fig2:sub1}) and (\ref{fig2:sub2}), we see that stronger magnetic fields result in a closer orbit at low energy values for charged black holes with a global monopole.

Even a weak external magnetic field shows a complex relationship with particle energies and orbital stability. At lower energies, an increasing magnetic field allows closer, stable orbits, effectively pulling the ISCO inward. However, its effect diminishes as $E\to 1$, where the ISCO behavior converges regardless of magnetic field strength. Notably, the magnetic field imposes a limiting factor on the ISCO at very low energies, preventing its divergence to infinity and providing a stabilizing influence instead of allowing it to reach infinity.

In order to see the combined effect of the magnetic field and the global monopole on the ISCO, we now take $\eta\neq0$. In Fig. (\ref{B and b:sub1}), we observe the attractive and repulsive nature of the external magnetic field and the global monopole. While $B$ allows for tighter orbits at very low energy values, the monopole pushes these orbits outwards. However, at very low energy values, the magnetic field's effect dominates. We also see that the monopole causes a constant, outward shift of the orbits. Note its effect in Fig. (\ref{B and b:sub2}). We also notice that as $B$ increases, the shift due to the monopole decreases. We discarded the $E\to 1$ limit since nothing interesting happens, and the results are exactly as shown in Fig. (\ref{fig1:sub2}). The effect of the magnetic field becomes nonexistent in this limit.

\begin{figure*}[h]
	\centering
	\begin{subfigure}{.5\textwidth}
		\centering
		\includegraphics[width=.89\linewidth]{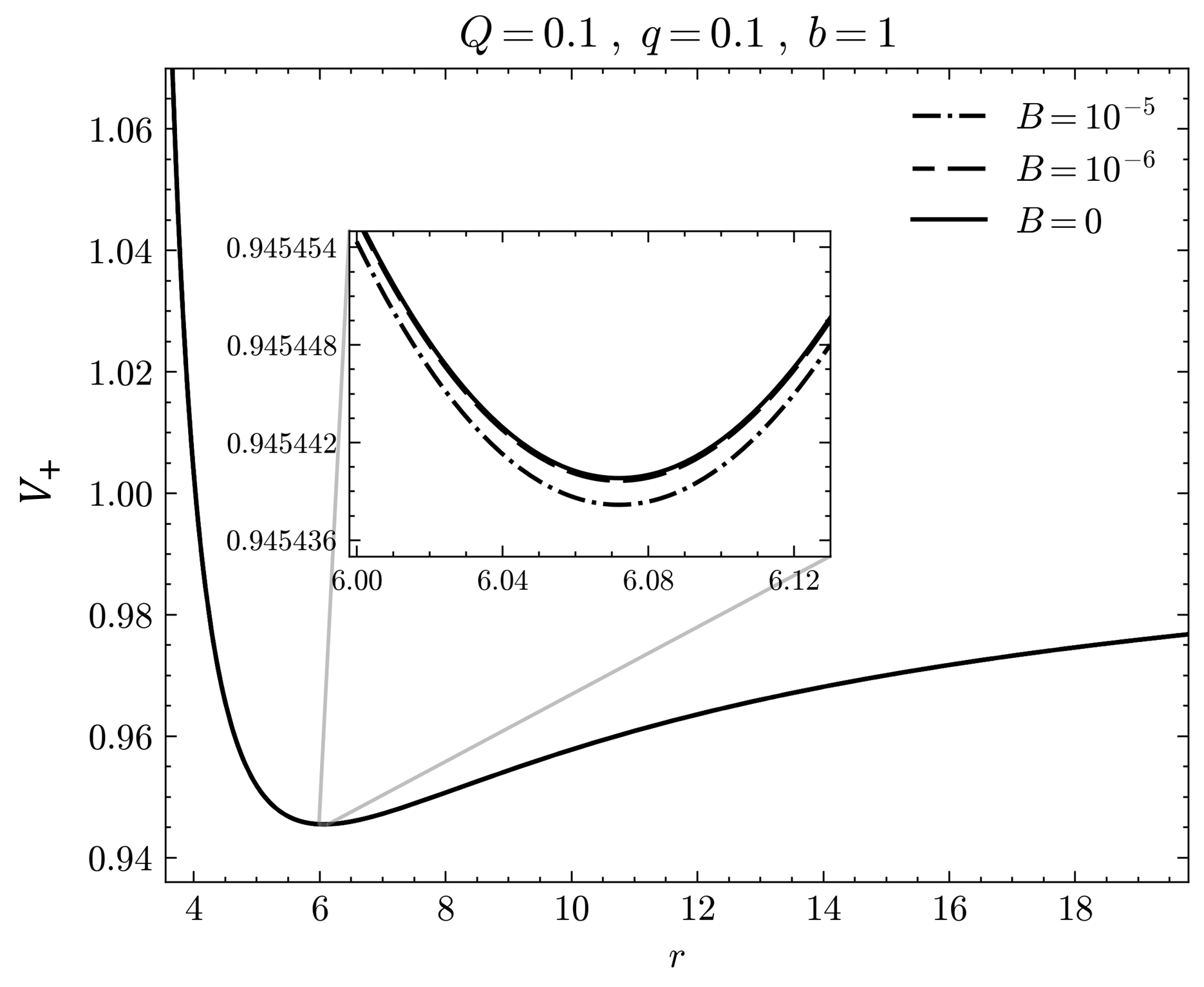}
		\caption{The effect of the external magnetic field, $B$, on $V_+$ %the effective potential  % for $\eta=0$ case
		}
		\label{fig3:sub1}
	\end{subfigure}%
	\begin{subfigure}{.5\textwidth}
		\centering
		\includegraphics[width=.89\linewidth]{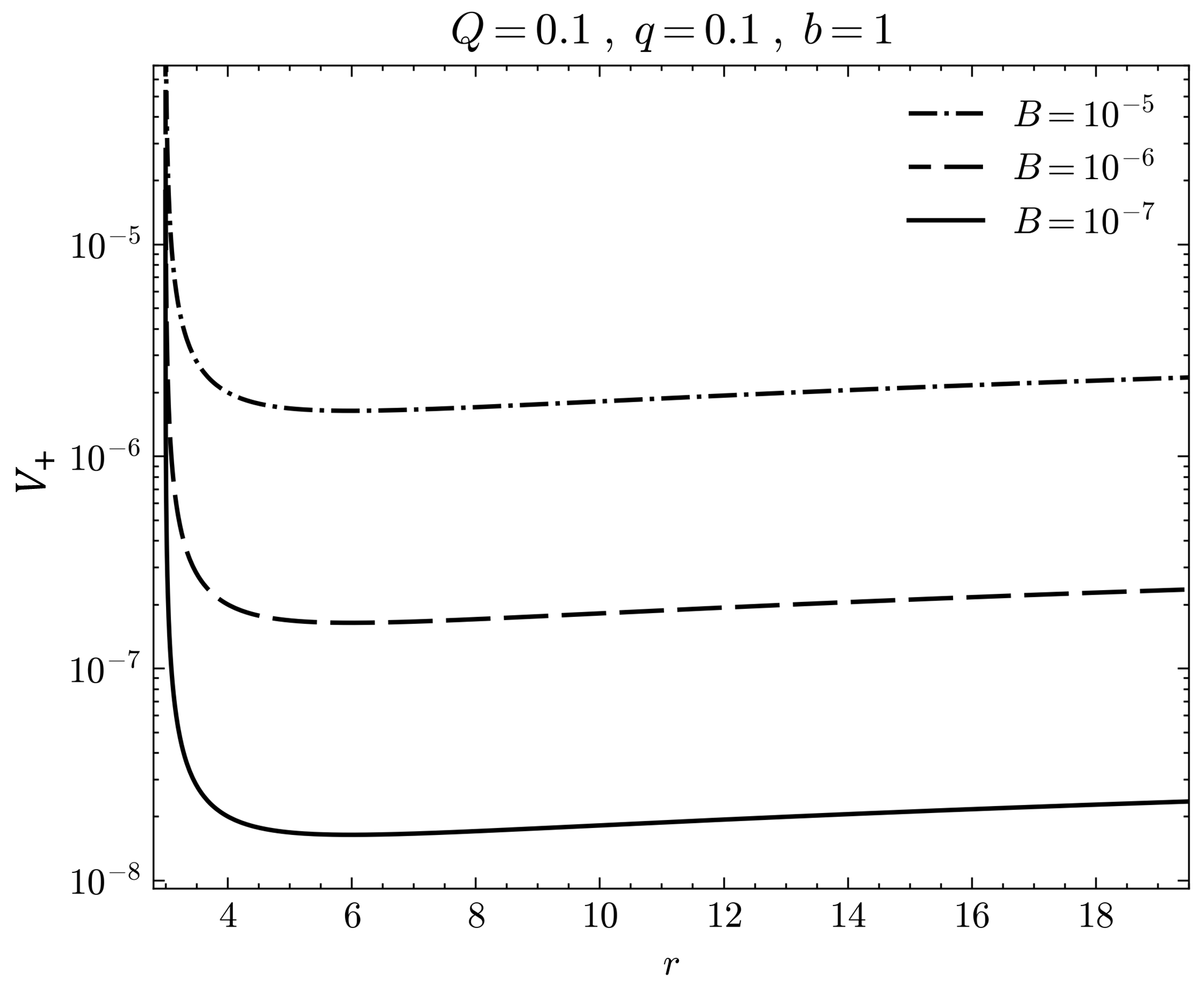}
		\caption{The graph of
			%for a given $B$ %magnetic field 
			%compared to the case with $B_=$ 
			%no magnetic field,  
			$\Delta V_+=V_+(B=0)-V_+(B)$ % is considered for different values of $B_i$ 
			%	in the $\eta = 0$ case
		}
		\label{fig3:sub2}
	\end{subfigure}
	\caption{Fig. (\ref{fig3:sub1}) is drawn for different values of the external magnetic field to focus on the behavior of the effective potential. Increasing the magnetic field does not change the location of the minimum of the potential, but it affects its strength by pushing the potential curve downwards, as can be seen by the zoomed-in part of the graph. Fig. (\ref{fig3:sub2}) concentrates on the differences in the effective potential when different values of  $B$  are present, starting from the $B=0$ case. Since the effect of $B$ is too small to show in the graph over its full range, we have preferred to plot the difference compared to the case where no external magnetic field is present. Note that the vertical axis is logarithmic in this figure. 
	}
	\label{fig:test2}
\end{figure*}

\begin{figure}[ht]
	\centering
	%	\begin{subfigure}{.5\textwidth}
		\centering
		\includegraphics[width=.59\linewidth]{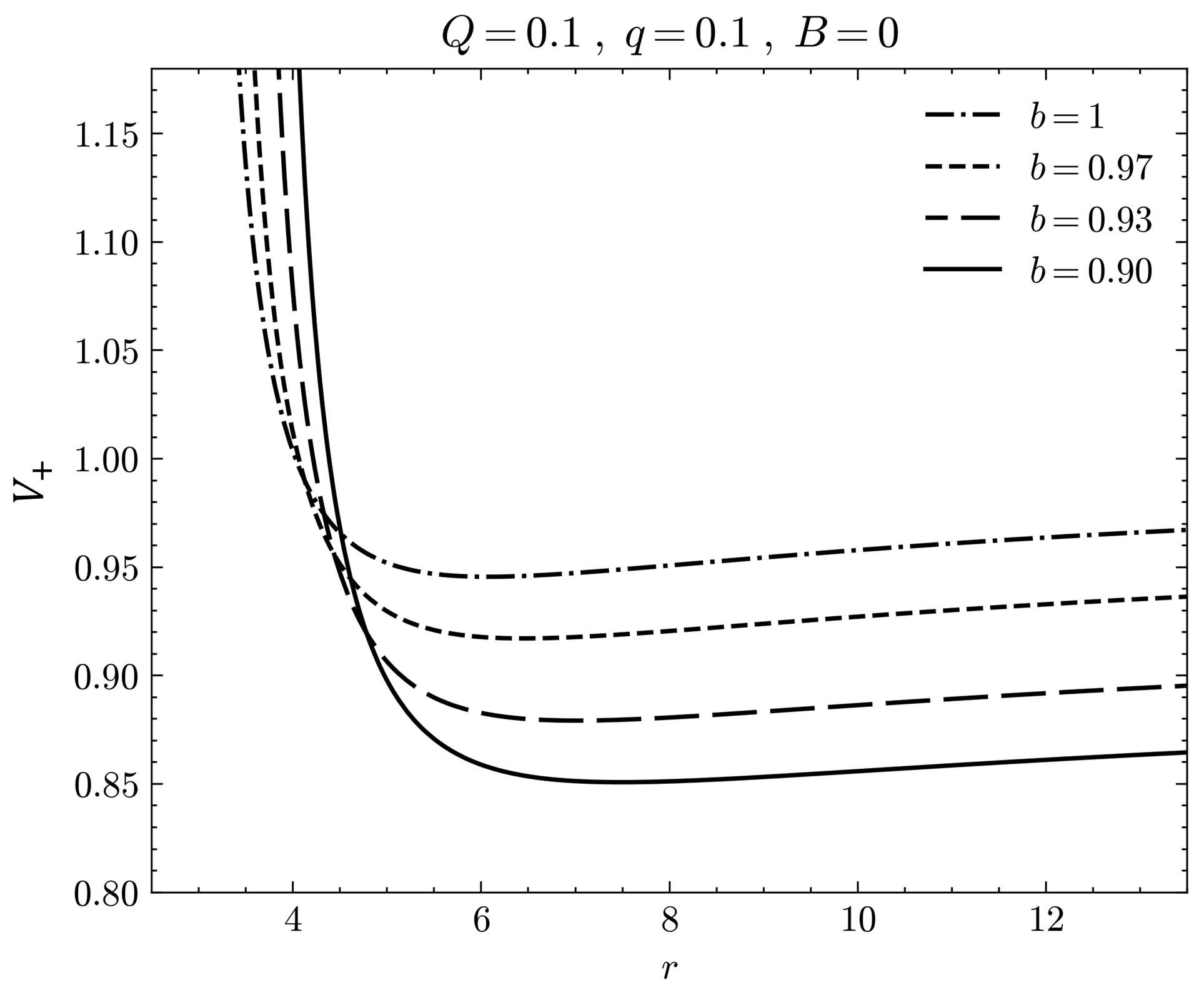}
		\caption{The graph is drawn for $B=0$, and for the different values of the monopole term $b$  to see its effect on the behavior of the effective potential, $V_+$. This results in a combination of both downward and rightward shifts in the potential curve.
			%Effect of $\eta$ on Effective Potential for $B=0$.
		}
		\label{fig4:sub1}
		%	\end{subfigure}%
	%	\begin{subfigure}{.5\textwidth}
		%		\centering
		%		\includegraphics[width=.89\linewidth]{Figure_4a_difference.png}
		%		\caption{The differences of Effective Potential for different values of $B$ for $\eta = 0$}
		%		\label{fig3:sub2}
		%	\end{subfigure}
	%	\caption{Fig.\ref{fig4:sub1} is drawn for the different values of the monopoleterm $b$  to see its effect on the behavior of the effective potential, $V+$, which results in a combination of both downward and rightward shifts on the potential
		%		curve.
		%	}
	%	\label{fig:test2a}
\end{figure}

\section{EFFECTIVE POTENTIAL ANALYSIS FOR THE ISCO}

	To paint the full picture of the behavior of charged test particles in RN spacetime with a global monopole under an external magnetic field, we may need to analyze the effective potential described by Eq. (\ref{expanded potentil}). We primarily focus on the behavior of the innermost stable circular orbit. Therefore, the potential must satisfy the following two conditions: The first is that we should have circular orbits, namely

\begin{figure*}[h]
	\centering
	\begin{subfigure}{.5\textwidth}
		\centering
		\includegraphics[width=.9\linewidth]{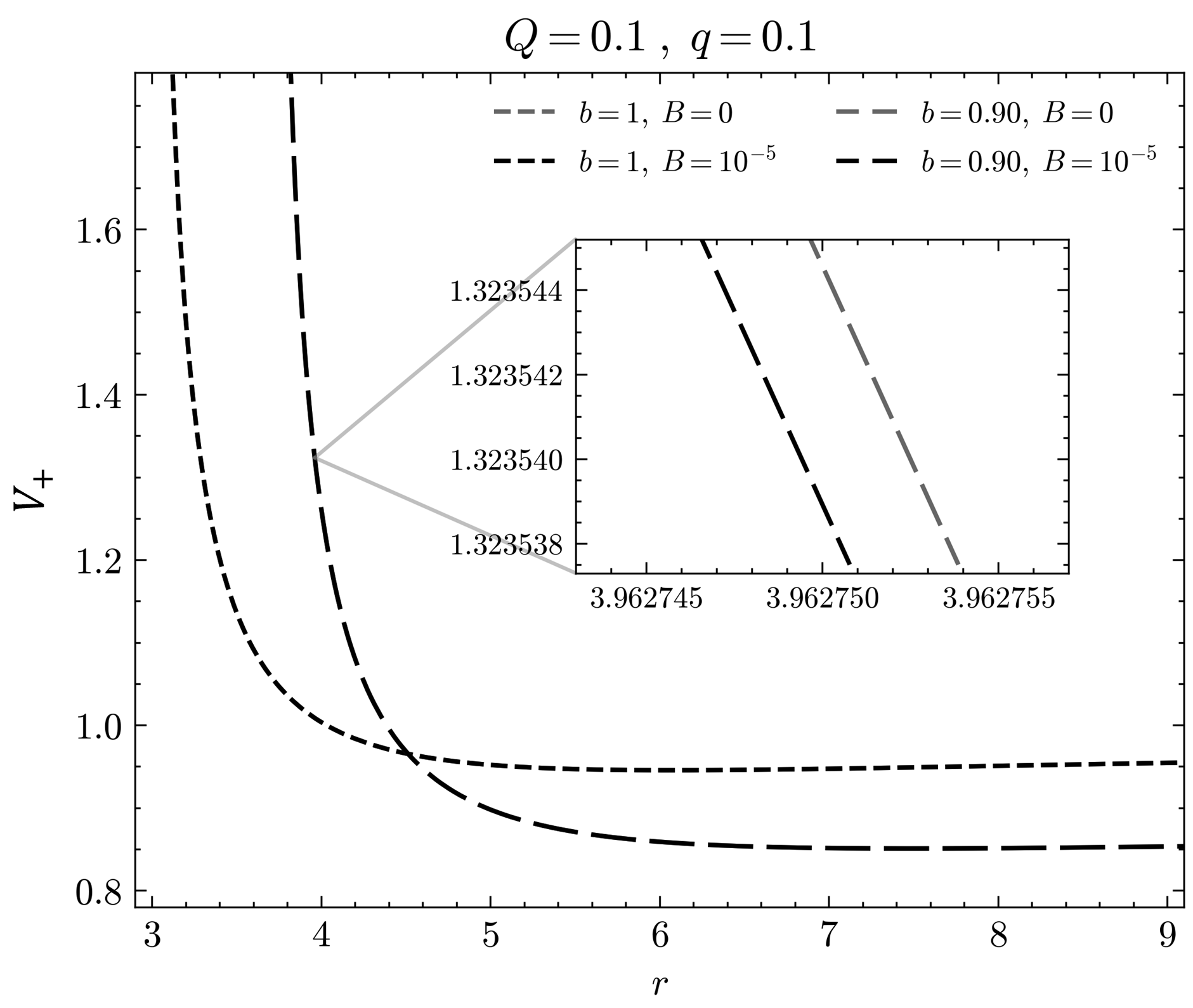}
		\caption{Effects of $B$ and $b$ on the effective potential %\\
			 for smaller $r$}
		\label{fig5:sub1}
	\end{subfigure}%
	\begin{subfigure}{.5\textwidth}
		\centering
		\includegraphics[width=.9\linewidth]{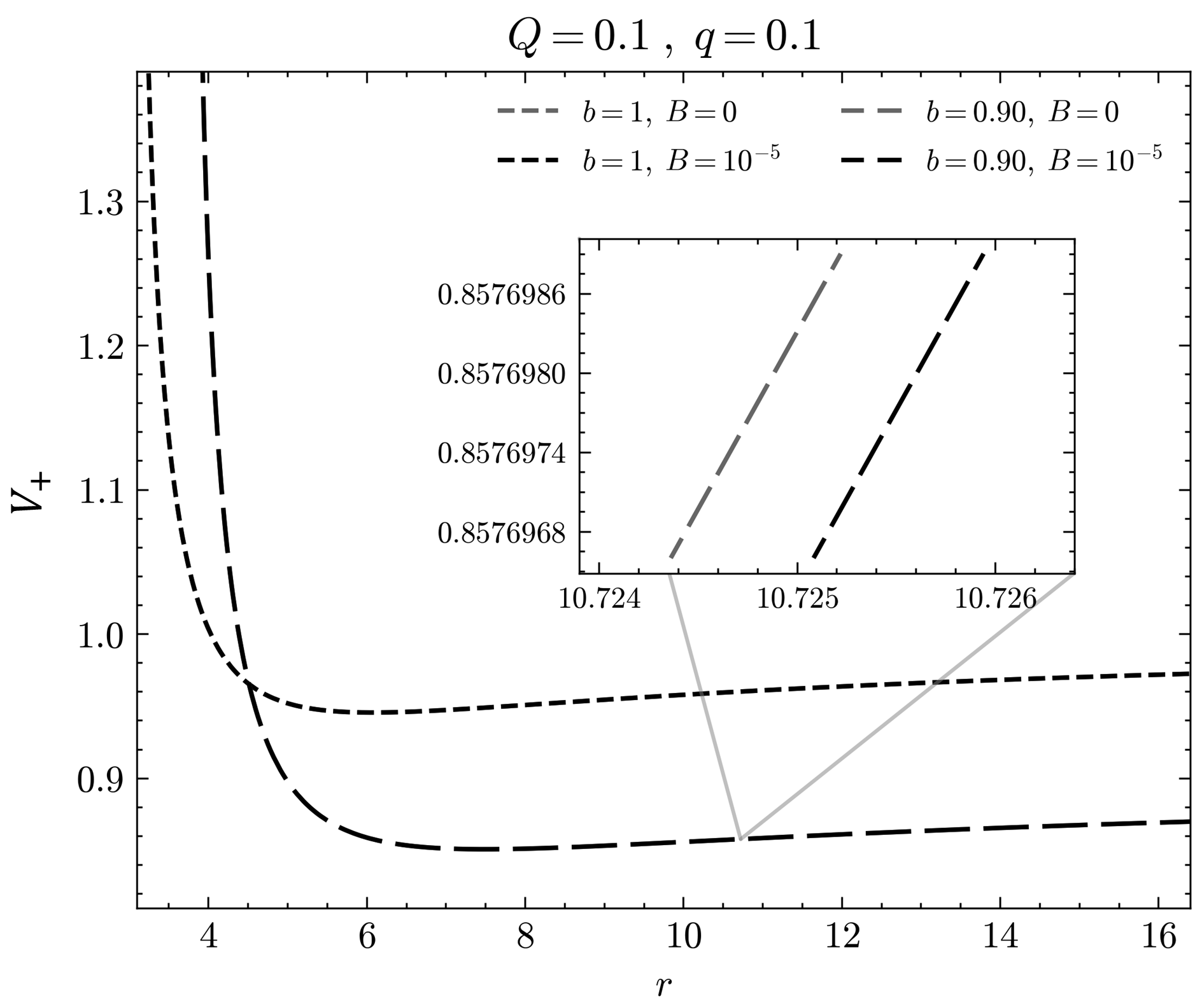}
		\caption{Effects of $B$ and $b$ on the effective potential %\\
			 for larger $r$}
		\label{fig5:sub2}
	\end{subfigure}
	\caption{Fig. (\ref{fig5:sub1}) focuses on the combined effect of the monopole and magnetic field on the regions closer to the horizon. We see that the general behavior is dominated by the global monopole. The effect of the magnetic field is too small and can be seen only if we zoom into the graphs at the given point.    Fig. (\ref{fig5:sub2}) focuses on such behavior for larger $r$ values, and both the global monopole and magnetic field show their effects on the effective potential, similar to smaller $r$ values.
	}
	\label{fig:test3}
\end{figure*}

\begin{eqnarray}
	\dot{r}^2 = 0,
\end{eqnarray}
in which, from Eq. (\ref{r dot pot}), we have the expression: 
\begin{eqnarray}
	(E-V_{+})(E-V_{-}) = 0.
\end{eqnarray}
We choose to analyze only $V_{+}$ as in \citep{article} to discuss behavior outside the event horizon.
The second condition is
\begin{eqnarray}\label{pot derivative 1}
	\frac{dV_{+}}{dr} = 0.
\end{eqnarray}
We start by solving (\ref{pot derivative 1}) for $L$, which allows us to focus on the effects of the magnetic field and the global monopole on the potential.

Fig. (\ref{fig3:sub1}) shows the behavior of the effective potential as a function of $r$ under the influence of an external magnetic field. One of the key observations is that under the influence of an external magnetic field, there is virtually no change to the point at $r$ where $V_+$ is at its minimum (ISCO); rather, we only observe a downward shift of the potential curve as $B$ increases, as the zoomed region of (\ref{fig3:sub1}) shows. This indicates that an external weak magnetic field will allow the ISCO at lower energies of orbiting particles. Fig. (\ref{fig4:sub1}) shows the effect of the global monopole on the effective potential. Similarly to the magnetic field, the global monopole causes a notable downward shift in the energy of the effective potential. Furthermore, as $b$ decreases, the potential curve moves to the right. This rightward shift indicates that the presence of the global monopole forces larger orbits. Thus, the monopole affects both the energy required for stable orbits and their location: it lowers the energy and pushes the orbits outward. This behavior of the potential well would make it easier for a particle to escape. This effect is  readily presented  in Fig. (\ref{fig1:sub2}) and  was interpreted as a repulsive effect of the global monopole.

Fig.s (\ref{fig5:sub1}) and (\ref{fig5:sub2}) illustrate the combined influence of the external magnetic field and the global monopole on the effective potential. Fig. (\ref{fig5:sub1}) focuses on the region close to the horizon, where $V\to \infty$. We see that as $r$ approaches the horizon, the effect of the magnetic field on the effective potential becomes more and more negligible. As we get closer to the horizon, the effect of the global monopole remains the same. Instead, we see Equation (\ref{horizon}) in action, where the monopole shifts the horizon inward. On the other hand, Fig. (\ref{fig5:sub2}) shows the potential at distances away from the horizon. Here, we see that as $r$ changes, the effect of both the global monopole and the external magnetic field remains constant. Other than that, the observations made for Fig. (\ref{fig3:sub1}) and Fig. (\ref{fig3:sub2}) stand true.

\section{CONCLUSION}

In this paper, we examined the dynamics of charged particles in the spacetime of a global monopole swallowed by a Reissner-Nordstr\"om black hole under the influence of a homogeneous external weak magnetic field. Our focus was on analyzing the behavior of the innermost stable circular orbit (ISCO) under various conditions. We generalized previous literature on the ISCO of black holes in a test magnetic field to have a nontrivial electrical charge and global monopole term. We determined the conditions required to have a constant and homogeneous magnetic field along an axis of symmetry around such a black hole. We have seen that if the electric charge of the black hole and the global monopole term are at least an order of magnitude smaller than the mass of the black hole, then we can have such a configuration. Using this approximation, we were able to deduce the necessary equations to have an ISCO around a Reissner-Nordstr\"om black hole with a global monopole embedded in a constant, homogeneous magnetic field, which has required a numerical approach due to the complexity of the resulting equations compared to neutral black holes in a similar setting.

In our numerical analysis, which was presented via several graphs, we observed that the presence of a global monopole significantly affects the behavior of the ISCO. As the monopole charge $\eta$ increases, the ISCO shifts outward, indicating a repulsive effect. This repulsive nature of the global monopole aligns with previous studies on the deflection angles of photons in \citep{Harari:1990cz,haluk}. The introduction of a weak external magnetic field shows a complex relationship between particle energies and orbital stability. At lower energies, stronger magnetic fields allow for closer, stable orbits, effectively pulling the ISCO inward. However, this effect diminishes as the particle energies approach $E\to1$, where the ISCO behavior converges regardless of the magnetic field strength. Notably, the magnetic field imposes a limiting factor on the ISCO at very low energies, preventing its divergence to infinity and providing a stabilizing influence instead of allowing it to recede to infinity.
The analysis of the effective potential shows that the monopole charge shifts the potential well outward and downward. This shift indicates larger orbital radii and lower energy requirements for stable orbits with a stronger monopole charge. The external magnetic field shifts the potential downward, reducing the energy that particles need to  orbit the ISCO for the same radius compared to when no external magnetic field is present. In other words,  particles may have closer ISCO orbits for the same energy in the presence of a magnetic field, and these results generalize similar observations for neutral black holes to nontrivially charged black holes.

One might expect that the present work can be easily expanded to more general cases; for example, a global monopole swallowed by a rotating, charged black hole and/or black holes with global monopoles in strong magnetic fields. Unfortunately, the lack of exact solutions corresponding to global monopoles swallowed by rotating, charged, or even neutral \citep{HalukSecuk:2020oos} black holes, with or without weak or strong magnetic fields, makes these generalizations impossible at the present time. Hence, the focus might be on obtaining such solutions in the presence of global monopoles for further studies that generalize the present discussion.

\section*{Acknowledgements}
We thank the anonymous referees for insightful comments about our manuscript.
M. H. S. and O. D. are supported by the Marmara University Scientific Research Projects Committee (Project Code: FDK-2021-10432).

%\nocite{*}
\bibliography{ref}

%apsrev4-2.bst 2019-01-14 (MD) hand-edited version of apsrev4-1.bst
%Control: key (0)
%Control: author (72) initials jnrlst
%Control: editor formatted (1) identically to author
%Control: production of article title (-1) disabled
%Control: page (0) single
%Control: year (1) truncated
%Control: production of eprint (0) enabled
\begin{thebibliography}{44}%
\makeatletter
\providecommand \@ifxundefined [1]{%
 \@ifx{#1\undefined}
}%
\providecommand \@ifnum [1]{%
 \ifnum #1\expandafter \@firstoftwo
 \else \expandafter \@secondoftwo
 \fi
}%
\providecommand \@ifx [1]{%
 \ifx #1\expandafter \@firstoftwo
 \else \expandafter \@secondoftwo
 \fi
}%
\providecommand \natexlab [1]{#1}%
\providecommand \enquote  [1]{``#1''}%
\providecommand \bibnamefont  [1]{#1}%
\providecommand \bibfnamefont [1]{#1}%
\providecommand \citenamefont [1]{#1}%
\providecommand \href@noop [0]{\@secondoftwo}%
\providecommand \href [0]{\begingroup \@sanitize@url \@href}%
\providecommand \@href[1]{\@@startlink{#1}\@@href}%
\providecommand \@@href[1]{\endgroup#1\@@endlink}%
\providecommand \@sanitize@url [0]{\catcode `\\12\catcode `\$12\catcode
  `\&12\catcode `\#12\catcode `\^12\catcode `\_12\catcode `\%12\relax}%
\providecommand \@@startlink[1]{}%
\providecommand \@@endlink[0]{}%
\providecommand \url  [0]{\begingroup\@sanitize@url \@url }%
\providecommand \@url [1]{\endgroup\@href {#1}{\urlprefix }}%
\providecommand \urlprefix  [0]{URL }%
\providecommand \Eprint [0]{\href }%
\providecommand \doibase [0]{https://doi.org/}%
\providecommand \selectlanguage [0]{\@gobble}%
\providecommand \bibinfo  [0]{\@secondoftwo}%
\providecommand \bibfield  [0]{\@secondoftwo}%
\providecommand \translation [1]{[#1]}%
\providecommand \BibitemOpen [0]{}%
\providecommand \bibitemStop [0]{}%
\providecommand \bibitemNoStop [0]{.\EOS\space}%
\providecommand \EOS [0]{\spacefactor3000\relax}%
\providecommand \BibitemShut  [1]{\csname bibitem#1\endcsname}%
\let\auto@bib@innerbib\@empty
%</preamble>
\bibitem [{\citenamefont {Frolov}\ and\ \citenamefont
  {Novikov}(1998)}]{Frolov:1998wf}%
  \BibitemOpen
  \bibinfo {editor} {\bibfnamefont {V.~P.}\ \bibnamefont {Frolov}}\ and\
  \bibinfo {editor} {\bibfnamefont {I.~D.}\ \bibnamefont {Novikov}},\ eds.,\
  \href {https://doi.org/10.1007/978-94-011-5139-9} {\emph {\bibinfo {title}
  {{Black hole physics: Basic concepts and new developments}}}}\ (\bibinfo
  {year} {1998})\BibitemShut {NoStop}%
\bibitem [{\citenamefont {Misner}\ \emph {et~al.}(1973)\citenamefont {Misner},
  \citenamefont {Thorne},\ and\ \citenamefont {Wheeler}}]{Misner:1973prb}%
  \BibitemOpen
  \bibfield  {author} {\bibinfo {author} {\bibfnamefont {C.~W.}\ \bibnamefont
  {Misner}}, \bibinfo {author} {\bibfnamefont {K.~S.}\ \bibnamefont {Thorne}},\
  and\ \bibinfo {author} {\bibfnamefont {J.~A.}\ \bibnamefont {Wheeler}},\
  }\href@noop {} {\emph {\bibinfo {title} {{Gravitation}}}}\ (\bibinfo
  {publisher} {W. H. Freeman},\ \bibinfo {address} {San Francisco},\ \bibinfo
  {year} {1973})\ \bibinfo {note} {pg. 901}\BibitemShut {NoStop}%
\bibitem [{\citenamefont {Shakura}\ and\ \citenamefont
  {Sunyaev}(1973)}]{Shakura:1972te}%
  \BibitemOpen
  \bibfield  {author} {\bibinfo {author} {\bibfnamefont {N.~I.}\ \bibnamefont
  {Shakura}}\ and\ \bibinfo {author} {\bibfnamefont {R.~A.}\ \bibnamefont
  {Sunyaev}},\ }\href@noop {} {\bibfield  {journal} {\bibinfo  {journal}
  {Astron. Astrophys.}\ }\textbf {\bibinfo {volume} {24}},\ \bibinfo {pages}
  {337} (\bibinfo {year} {1973})}\BibitemShut {NoStop}%
\bibitem [{\citenamefont {Abramowicz}\ and\ \citenamefont
  {Fragile}(2013)}]{Abramowicz:2011xu}%
  \BibitemOpen
  \bibfield  {author} {\bibinfo {author} {\bibfnamefont {M.~A.}\ \bibnamefont
  {Abramowicz}}\ and\ \bibinfo {author} {\bibfnamefont {P.~C.}\ \bibnamefont
  {Fragile}},\ }\href {https://doi.org/10.12942/lrr-2013-1} {\bibfield
  {journal} {\bibinfo  {journal} {Living Rev. Rel.}\ }\textbf {\bibinfo
  {volume} {16}},\ \bibinfo {pages} {1} (\bibinfo {year} {2013})},\ \Eprint
  {https://arxiv.org/abs/1104.5499} {arXiv:1104.5499 [astro-ph.HE]}
  \BibitemShut {NoStop}%
\bibitem [{\citenamefont {M.~Abramowicz}\ and\ \citenamefont
  {Sikora}(1978)}]{Abramowicz:1978}%
  \BibitemOpen
  \bibfield  {author} {\bibinfo {author} {\bibfnamefont {M.~J.}\ \bibnamefont
  {M.~Abramowicz}}\ and\ \bibinfo {author} {\bibfnamefont {M.}~\bibnamefont
  {Sikora}},\ }\href@noop {} {\bibfield  {journal} {\bibinfo  {journal} {{
  Astron. Astrophys.}}\ }\textbf {\bibinfo {volume} {63}},\ \bibinfo {pages}
  {221} (\bibinfo {year} {1978})}\BibitemShut {NoStop}%
\bibitem [{\citenamefont {Zakharov}\ \emph {et~al.}(2003)\citenamefont
  {Zakharov}, \citenamefont {Kardashev}, \citenamefont {Lukash},\ and\
  \citenamefont {Repin}}]{Zakharov_2003}%
  \BibitemOpen
  \bibfield  {author} {\bibinfo {author} {\bibfnamefont {A.~F.}\ \bibnamefont
  {Zakharov}}, \bibinfo {author} {\bibfnamefont {N.~S.}\ \bibnamefont
  {Kardashev}}, \bibinfo {author} {\bibfnamefont {V.~N.}\ \bibnamefont
  {Lukash}},\ and\ \bibinfo {author} {\bibfnamefont {S.~V.}\ \bibnamefont
  {Repin}},\ }\href {https://doi.org/10.1046/j.1365-8711.2003.06638.x}
  {\bibfield  {journal} {\bibinfo  {journal} {Monthly Notices of the Royal
  Astronomical Society}\ }\textbf {\bibinfo {volume} {342}},\ \bibinfo {pages}
  {1325–1333} (\bibinfo {year} {2003})}\BibitemShut {NoStop}%
\bibitem [{\citenamefont {Barriola}\ and\ \citenamefont
  {Vilenkin}(1989)}]{Barriola:1989hx}%
  \BibitemOpen
  \bibfield  {author} {\bibinfo {author} {\bibfnamefont {M.}~\bibnamefont
  {Barriola}}\ and\ \bibinfo {author} {\bibfnamefont {A.}~\bibnamefont
  {Vilenkin}},\ }\href {https://doi.org/10.1103/PhysRevLett.63.341} {\bibfield
  {journal} {\bibinfo  {journal} {Phys. Rev. Lett.}\ }\textbf {\bibinfo
  {volume} {63}},\ \bibinfo {pages} {341} (\bibinfo {year} {1989})}\BibitemShut
  {NoStop}%
\bibitem [{\citenamefont {Kibble}(1976)}]{Kibble:1976sj}%
  \BibitemOpen
  \bibfield  {author} {\bibinfo {author} {\bibfnamefont {T.~W.~B.}\
  \bibnamefont {Kibble}},\ }\href {https://doi.org/10.1088/0305-4470/9/8/029}
  {\bibfield  {journal} {\bibinfo  {journal} {J. Phys. A}\ }\textbf {\bibinfo
  {volume} {9}},\ \bibinfo {pages} {1387} (\bibinfo {year} {1976})}\BibitemShut
  {NoStop}%
\bibitem [{\citenamefont {Vilenkin}(1985)}]{Vilenkin:1984ib}%
  \BibitemOpen
  \bibfield  {author} {\bibinfo {author} {\bibfnamefont {A.}~\bibnamefont
  {Vilenkin}},\ }\href {https://doi.org/10.1016/0370-1573(85)90033-X}
  {\bibfield  {journal} {\bibinfo  {journal} {Phys. Rept.}\ }\textbf {\bibinfo
  {volume} {121}},\ \bibinfo {pages} {263} (\bibinfo {year}
  {1985})}\BibitemShut {NoStop}%
\bibitem [{\citenamefont {Harari}\ and\ \citenamefont
  {Lousto}(1990)}]{Harari:1990cz}%
  \BibitemOpen
  \bibfield  {author} {\bibinfo {author} {\bibfnamefont {D.}~\bibnamefont
  {Harari}}\ and\ \bibinfo {author} {\bibfnamefont {C.}~\bibnamefont
  {Lousto}},\ }\href {https://doi.org/10.1103/PhysRevD.42.2626} {\bibfield
  {journal} {\bibinfo  {journal} {Phys. Rev. D}\ }\textbf {\bibinfo {volume}
  {42}},\ \bibinfo {pages} {2626} (\bibinfo {year} {1990})}\BibitemShut
  {NoStop}%
\bibitem [{\citenamefont {Bennett}\ and\ \citenamefont
  {Rhie}(1990)}]{Bennett:1990xy}%
  \BibitemOpen
  \bibfield  {author} {\bibinfo {author} {\bibfnamefont {D.~P.}\ \bibnamefont
  {Bennett}}\ and\ \bibinfo {author} {\bibfnamefont {S.~H.}\ \bibnamefont
  {Rhie}},\ }\href {https://doi.org/10.1103/PhysRevLett.65.1709} {\bibfield
  {journal} {\bibinfo  {journal} {Phys. Rev. Lett.}\ }\textbf {\bibinfo
  {volume} {65}},\ \bibinfo {pages} {1709} (\bibinfo {year}
  {1990})}\BibitemShut {NoStop}%
\bibitem [{\citenamefont {Pogosian}\ \emph {et~al.}(2003)\citenamefont
  {Pogosian}, \citenamefont {Tye}, \citenamefont {Wasserman},\ and\
  \citenamefont {Wyman}}]{Pogosian:2003mz}%
  \BibitemOpen
  \bibfield  {author} {\bibinfo {author} {\bibfnamefont {L.}~\bibnamefont
  {Pogosian}}, \bibinfo {author} {\bibfnamefont {S.~H.~H.}\ \bibnamefont
  {Tye}}, \bibinfo {author} {\bibfnamefont {I.}~\bibnamefont {Wasserman}},\
  and\ \bibinfo {author} {\bibfnamefont {M.}~\bibnamefont {Wyman}},\ }\href
  {https://doi.org/10.1103/PhysRevD.68.023506} {\bibfield  {journal} {\bibinfo
  {journal} {Phys. Rev. D}\ }\textbf {\bibinfo {volume} {68}},\ \bibinfo
  {pages} {023506} (\bibinfo {year} {2003})},\ \bibinfo {note} {[Erratum:
  Phys.Rev.D 73, 089904 (2006)]},\ \Eprint
  {https://arxiv.org/abs/hep-th/0304188} {arXiv:hep-th/0304188} \BibitemShut
  {NoStop}%
\bibitem [{\citenamefont {Bevis}\ \emph {et~al.}(2007)\citenamefont {Bevis},
  \citenamefont {Hindmarsh}, \citenamefont {Kunz},\ and\ \citenamefont
  {Urrestilla}}]{Bevis:2006mj}%
  \BibitemOpen
  \bibfield  {author} {\bibinfo {author} {\bibfnamefont {N.}~\bibnamefont
  {Bevis}}, \bibinfo {author} {\bibfnamefont {M.}~\bibnamefont {Hindmarsh}},
  \bibinfo {author} {\bibfnamefont {M.}~\bibnamefont {Kunz}},\ and\ \bibinfo
  {author} {\bibfnamefont {J.}~\bibnamefont {Urrestilla}},\ }\href
  {https://doi.org/10.1103/PhysRevD.75.065015} {\bibfield  {journal} {\bibinfo
  {journal} {Phys. Rev. D}\ }\textbf {\bibinfo {volume} {75}},\ \bibinfo
  {pages} {065015} (\bibinfo {year} {2007})},\ \Eprint
  {https://arxiv.org/abs/astro-ph/0605018} {arXiv:astro-ph/0605018}
  \BibitemShut {NoStop}%
\bibitem [{\citenamefont {Guendelman}\ and\ \citenamefont
  {Rabinowitz}(1991)}]{PhysRevD.44.3152}%
  \BibitemOpen
  \bibfield  {author} {\bibinfo {author} {\bibfnamefont {E.~I.}\ \bibnamefont
  {Guendelman}}\ and\ \bibinfo {author} {\bibfnamefont {A.}~\bibnamefont
  {Rabinowitz}},\ }\href {https://doi.org/10.1103/PhysRevD.44.3152} {\bibfield
  {journal} {\bibinfo  {journal} {Phys. Rev. D}\ }\textbf {\bibinfo {volume}
  {44}},\ \bibinfo {pages} {3152} (\bibinfo {year} {1991})}\BibitemShut
  {NoStop}%
\bibitem [{\citenamefont {Li}\ and\ \citenamefont
  {Hao}(2002)}]{PhysRevD.66.107701}%
  \BibitemOpen
  \bibfield  {author} {\bibinfo {author} {\bibfnamefont {X.-z.}\ \bibnamefont
  {Li}}\ and\ \bibinfo {author} {\bibfnamefont {J.-g.}\ \bibnamefont {Hao}},\
  }\href {https://doi.org/10.1103/PhysRevD.66.107701} {\bibfield  {journal}
  {\bibinfo  {journal} {Phys. Rev. D}\ }\textbf {\bibinfo {volume} {66}},\
  \bibinfo {pages} {107701} (\bibinfo {year} {2002})}\BibitemShut {NoStop}%
\bibitem [{\citenamefont {Bertrand}\ \emph {et~al.}(2003)\citenamefont
  {Bertrand}, \citenamefont {Brihaye},\ and\ \citenamefont
  {Hartmann}}]{Bertrand_2003}%
  \BibitemOpen
  \bibfield  {author} {\bibinfo {author} {\bibfnamefont {B.}~\bibnamefont
  {Bertrand}}, \bibinfo {author} {\bibfnamefont {Y.}~\bibnamefont {Brihaye}},\
  and\ \bibinfo {author} {\bibfnamefont {B.}~\bibnamefont {Hartmann}},\ }\href
  {https://doi.org/10.1088/0264-9381/20/20/312} {\bibfield  {journal} {\bibinfo
   {journal} {Classical and Quantum Gravity}\ }\textbf {\bibinfo {volume}
  {20}},\ \bibinfo {pages} {4495} (\bibinfo {year} {2003})}\BibitemShut
  {NoStop}%
\bibitem [{\citenamefont {Delice}(2003)}]{OzgurDelice_2003}%
  \BibitemOpen
  \bibfield  {author} {\bibinfo {author} {\bibfnamefont {O.}~\bibnamefont
  {Delice}},\ }\href {https://doi.org/10.1088/1126-6708/2003/11/058} {\bibfield
   {journal} {\bibinfo  {journal} {Journal of High Energy Physics}\ }\textbf
  {\bibinfo {volume} {2003}},\ \bibinfo {pages} {058} (\bibinfo {year}
  {2003})}\BibitemShut {NoStop}%
\bibitem [{\citenamefont {Se\c{c}uk}\ and\ \citenamefont
  {Delice}(2020{\natexlab{a}})}]{Se_uk_2020}%
  \BibitemOpen
  \bibfield  {author} {\bibinfo {author} {\bibfnamefont {M.~H.}\ \bibnamefont
  {Se\c{c}uk}}\ and\ \bibinfo {author} {\bibfnamefont {O.}~\bibnamefont
  {Delice}},\ }\bibfield  {journal} {\bibinfo  {journal} {The European Physical
  Journal C}\ }\textbf {\bibinfo {volume} {80}},\ \href
  {https://doi.org/10.1140/epjc/s10052-020-7988-5}
  {10.1140/epjc/s10052-020-7988-5} (\bibinfo {year}
  {2020}{\natexlab{a}})\BibitemShut {NoStop}%
\bibitem [{\citenamefont {Se\c{c}uk}\ and\ \citenamefont
  {Delice}(2020{\natexlab{b}})}]{haluk}%
  \BibitemOpen
  \bibfield  {author} {\bibinfo {author} {\bibfnamefont {M.~H.}\ \bibnamefont
  {Se\c{c}uk}}\ and\ \bibinfo {author} {\bibfnamefont {O.}~\bibnamefont
  {Delice}},\ }\bibfield  {journal} {\bibinfo  {journal} {The European Physical
  Journal Plus}\ }\textbf {\bibinfo {volume} {135}},\ \href
  {https://doi.org/10.1140/epjp/s13360-020-00551-0}
  {10.1140/epjp/s13360-020-00551-0} (\bibinfo {year}
  {2020}{\natexlab{b}})\BibitemShut {NoStop}%
\bibitem [{\citenamefont {Schroven}\ and\ \citenamefont
  {Grunau}(2021)}]{Schroven_2021}%
  \BibitemOpen
  \bibfield  {author} {\bibinfo {author} {\bibfnamefont {K.}~\bibnamefont
  {Schroven}}\ and\ \bibinfo {author} {\bibfnamefont {S.}~\bibnamefont
  {Grunau}},\ }\bibfield  {journal} {\bibinfo  {journal} {Physical Review D}\
  }\textbf {\bibinfo {volume} {103}},\ \href
  {https://doi.org/10.1103/physrevd.103.024016} {10.1103/physrevd.103.024016}
  (\bibinfo {year} {2021})\BibitemShut {NoStop}%
\bibitem [{\citenamefont {Wald}(1974)}]{Wald:1974np}%
  \BibitemOpen
  \bibfield  {author} {\bibinfo {author} {\bibfnamefont {R.~M.}\ \bibnamefont
  {Wald}},\ }\href {https://doi.org/10.1103/PhysRevD.10.1680} {\bibfield
  {journal} {\bibinfo  {journal} {Phys. Rev. D}\ }\textbf {\bibinfo {volume}
  {10}},\ \bibinfo {pages} {1680} (\bibinfo {year} {1974})}\BibitemShut
  {NoStop}%
\bibitem [{\citenamefont {Prasanna}\ and\ \citenamefont
  {Vishveshwara}(1978)}]{Prasanna:1978vh}%
  \BibitemOpen
  \bibfield  {author} {\bibinfo {author} {\bibfnamefont {A.~R.}\ \bibnamefont
  {Prasanna}}\ and\ \bibinfo {author} {\bibfnamefont {C.~V.}\ \bibnamefont
  {Vishveshwara}},\ }\href {https://doi.org/10.1007/BF02848160} {\bibfield
  {journal} {\bibinfo  {journal} {Pramana}\ }\textbf {\bibinfo {volume} {11}},\
  \bibinfo {pages} {359} (\bibinfo {year} {1978})}\BibitemShut {NoStop}%
\bibitem [{\citenamefont {Frolov}\ and\ \citenamefont
  {Shoom}(2010)}]{Frolov:2010mi}%
  \BibitemOpen
  \bibfield  {author} {\bibinfo {author} {\bibfnamefont {V.~P.}\ \bibnamefont
  {Frolov}}\ and\ \bibinfo {author} {\bibfnamefont {A.~A.}\ \bibnamefont
  {Shoom}},\ }\href {https://doi.org/10.1103/PhysRevD.82.084034} {\bibfield
  {journal} {\bibinfo  {journal} {Phys. Rev. D}\ }\textbf {\bibinfo {volume}
  {82}},\ \bibinfo {pages} {084034} (\bibinfo {year} {2010})},\ \Eprint
  {https://arxiv.org/abs/1008.2985} {arXiv:1008.2985 [gr-qc]} \BibitemShut
  {NoStop}%
\bibitem [{\citenamefont {Aliev}\ and\ \citenamefont
  {Gal'tsov}(1989)}]{ANAliev_1989}%
  \BibitemOpen
  \bibfield  {author} {\bibinfo {author} {\bibfnamefont {A.~N.}\ \bibnamefont
  {Aliev}}\ and\ \bibinfo {author} {\bibfnamefont {D.~V.}\ \bibnamefont
  {Gal'tsov}},\ }\href {https://doi.org/10.1070/PU1989v032n01ABEH002677}
  {\bibfield  {journal} {\bibinfo  {journal} {Soviet Physics Uspekhi}\ }\textbf
  {\bibinfo {volume} {32}},\ \bibinfo {pages} {75} (\bibinfo {year}
  {1989})}\BibitemShut {NoStop}%
\bibitem [{\citenamefont {Aliev}\ and\ \citenamefont
  {\"Ozdemir}(2002)}]{AlievANandOzdemirN}%
  \BibitemOpen
  \bibfield  {author} {\bibinfo {author} {\bibfnamefont {A.~N.}\ \bibnamefont
  {Aliev}}\ and\ \bibinfo {author} {\bibfnamefont {N.}~\bibnamefont
  {\"Ozdemir}},\ }\href {https://doi.org/10.1046/j.1365-8711.2002.05727.x}
  {\bibfield  {journal} {\bibinfo  {journal} {Monthly Notices of the Royal
  Astronomical Society}\ }\textbf {\bibinfo {volume} {336}},\ \bibinfo {pages}
  {241} (\bibinfo {year} {2002})}\BibitemShut {NoStop}%
\bibitem [{\citenamefont {Baker}\ and\ \citenamefont
  {Frolov}(2023)}]{Baker:2023gdc}%
  \BibitemOpen
  \bibfield  {author} {\bibinfo {author} {\bibfnamefont {N.~P.}\ \bibnamefont
  {Baker}}\ and\ \bibinfo {author} {\bibfnamefont {V.~P.}\ \bibnamefont
  {Frolov}},\ }\href {https://doi.org/10.1103/PhysRevD.108.024045} {\bibfield
  {journal} {\bibinfo  {journal} {Phys. Rev. D}\ }\textbf {\bibinfo {volume}
  {108}},\ \bibinfo {pages} {024045} (\bibinfo {year} {2023})},\ \Eprint
  {https://arxiv.org/abs/2305.12591} {arXiv:2305.12591 [gr-qc]} \BibitemShut
  {NoStop}%
\bibitem [{\citenamefont {Hackstein}\ and\ \citenamefont
  {Hackmann}(2020)}]{Hackstein:2019msh}%
  \BibitemOpen
  \bibfield  {author} {\bibinfo {author} {\bibfnamefont {J.~P.}\ \bibnamefont
  {Hackstein}}\ and\ \bibinfo {author} {\bibfnamefont {E.}~\bibnamefont
  {Hackmann}},\ }\href {https://doi.org/10.1007/s10714-020-02675-1} {\bibfield
  {journal} {\bibinfo  {journal} {Gen. Rel. Grav.}\ }\textbf {\bibinfo {volume}
  {52}},\ \bibinfo {pages} {22} (\bibinfo {year} {2020})},\ \Eprint
  {https://arxiv.org/abs/1911.07645} {arXiv:1911.07645 [gr-qc]} \BibitemShut
  {NoStop}%
\bibitem [{\citenamefont {Al~Zahrani}(2021)}]{AlZahrani:2021jxe}%
  \BibitemOpen
  \bibfield  {author} {\bibinfo {author} {\bibfnamefont {A.~M.}\ \bibnamefont
  {Al~Zahrani}},\ }\href {https://doi.org/10.1103/PhysRevD.103.084008}
  {\bibfield  {journal} {\bibinfo  {journal} {Phys. Rev. D}\ }\textbf {\bibinfo
  {volume} {103}},\ \bibinfo {pages} {084008} (\bibinfo {year} {2021})},\
  \Eprint {https://arxiv.org/abs/2208.13956} {arXiv:2208.13956 [gr-qc]}
  \BibitemShut {NoStop}%
\bibitem [{\citenamefont {Al~Zahrani}(2014)}]{AlZahrani:2014dfi}%
  \BibitemOpen
  \bibfield  {author} {\bibinfo {author} {\bibfnamefont {A.~M.}\ \bibnamefont
  {Al~Zahrani}},\ }\href {https://doi.org/10.1103/PhysRevD.90.044012}
  {\bibfield  {journal} {\bibinfo  {journal} {Phys. Rev. D}\ }\textbf {\bibinfo
  {volume} {90}},\ \bibinfo {pages} {044012} (\bibinfo {year} {2014})},\
  \Eprint {https://arxiv.org/abs/1407.7069} {arXiv:1407.7069 [gr-qc]}
  \BibitemShut {NoStop}%
\bibitem [{\citenamefont {Al~Zahrani}(2022)}]{AlZahrani:2022fas}%
  \BibitemOpen
  \bibfield  {author} {\bibinfo {author} {\bibfnamefont {A.~M.}\ \bibnamefont
  {Al~Zahrani}},\ }\href {https://doi.org/10.3847/1538-4357/ac8cf0} {\bibfield
  {journal} {\bibinfo  {journal} {Astrophys. J.}\ }\textbf {\bibinfo {volume}
  {937}},\ \bibinfo {pages} {50} (\bibinfo {year} {2022})},\ \Eprint
  {https://arxiv.org/abs/2209.00144} {arXiv:2209.00144 [gr-qc]} \BibitemShut
  {NoStop}%
\bibitem [{\citenamefont {Schroven}\ \emph {et~al.}(2017)\citenamefont
  {Schroven}, \citenamefont {Hackmann},\ and\ \citenamefont
  {L\"ammerzahl}}]{PhysRevD.96.063015}%
  \BibitemOpen
  \bibfield  {author} {\bibinfo {author} {\bibfnamefont {K.}~\bibnamefont
  {Schroven}}, \bibinfo {author} {\bibfnamefont {E.}~\bibnamefont {Hackmann}},\
  and\ \bibinfo {author} {\bibfnamefont {C.}~\bibnamefont {L\"ammerzahl}},\
  }\href {https://doi.org/10.1103/PhysRevD.96.063015} {\bibfield  {journal}
  {\bibinfo  {journal} {Phys. Rev. D}\ }\textbf {\bibinfo {volume} {96}},\
  \bibinfo {pages} {063015} (\bibinfo {year} {2017})}\BibitemShut {NoStop}%
\bibitem [{\citenamefont {Piotrovich}\ \emph {et~al.}(2010)\citenamefont
  {Piotrovich}, \citenamefont {Silant'ev}, \citenamefont {Gnedin},\ and\
  \citenamefont {Natsvlishvili}}]{Piotrovich:2010aq}%
  \BibitemOpen
  \bibfield  {author} {\bibinfo {author} {\bibfnamefont {M.~Y.}\ \bibnamefont
  {Piotrovich}}, \bibinfo {author} {\bibfnamefont {N.~A.}\ \bibnamefont
  {Silant'ev}}, \bibinfo {author} {\bibfnamefont {Y.~N.}\ \bibnamefont
  {Gnedin}},\ and\ \bibinfo {author} {\bibfnamefont {T.~M.}\ \bibnamefont
  {Natsvlishvili}},\ }\href@noop {} {\bibfield  {journal} {\bibinfo  {journal}
  {arXiv}\ }\textbf {\bibinfo {volume} {1002.4948[astro-ph.CO]}} (\bibinfo
  {year} {2010})},\ \Eprint {https://arxiv.org/abs/1002.4948[astro-ph.CO]}
  {arXiv:1002.4948[astro-ph.CO] [astro-ph.CO]} \BibitemShut {NoStop}%
\bibitem [{\citenamefont {Melvin}(1965)}]{PhysRev.139.B225}%
  \BibitemOpen
  \bibfield  {author} {\bibinfo {author} {\bibfnamefont {M.~A.}\ \bibnamefont
  {Melvin}},\ }\href {https://doi.org/10.1103/PhysRev.139.B225} {\bibfield
  {journal} {\bibinfo  {journal} {Phys. Rev.}\ }\textbf {\bibinfo {volume}
  {139}},\ \bibinfo {pages} {B225} (\bibinfo {year} {1965})}\BibitemShut
  {NoStop}%
\bibitem [{\citenamefont {Ernst}(1976)}]{10.1063/1.522781}%
  \BibitemOpen
  \bibfield  {author} {\bibinfo {author} {\bibfnamefont {F.~J.}\ \bibnamefont
  {Ernst}},\ }\href {https://doi.org/10.1063/1.522781} {\bibfield  {journal}
  {\bibinfo  {journal} {Journal of Mathematical Physics}\ }\textbf {\bibinfo
  {volume} {17}},\ \bibinfo {pages} {54} (\bibinfo {year} {1976})}\BibitemShut
  {NoStop}%
\bibitem [{\citenamefont {Gibbons}\ \emph {et~al.}(2013)\citenamefont
  {Gibbons}, \citenamefont {Mujtaba},\ and\ \citenamefont
  {Pope}}]{Gibbons_2013}%
  \BibitemOpen
  \bibfield  {author} {\bibinfo {author} {\bibfnamefont {G.~W.}\ \bibnamefont
  {Gibbons}}, \bibinfo {author} {\bibfnamefont {A.~H.}\ \bibnamefont
  {Mujtaba}},\ and\ \bibinfo {author} {\bibfnamefont {C.~N.}\ \bibnamefont
  {Pope}},\ }\href {https://doi.org/10.1088/0264-9381/30/12/125008} {\bibfield
  {journal} {\bibinfo  {journal} {Classical and Quantum Gravity}\ }\textbf
  {\bibinfo {volume} {30}},\ \bibinfo {pages} {125008} (\bibinfo {year}
  {2013})}\BibitemShut {NoStop}%
\bibitem [{\citenamefont {Mandal}(2023)}]{Mandal:2023bgw}%
  \BibitemOpen
  \bibfield  {author} {\bibinfo {author} {\bibfnamefont {P.}~\bibnamefont
  {Mandal}},\ }\href {https://doi.org/10.30970/jps.27.4901} {\bibfield
  {journal} {\bibinfo  {journal} {J. Phys. Stud.}\ }\textbf {\bibinfo {volume}
  {27}},\ \bibinfo {pages} {4901} (\bibinfo {year} {2023})}\BibitemShut
  {NoStop}%
\bibitem [{\citenamefont {Majeed}\ \emph {et~al.}(2015)\citenamefont {Majeed},
  \citenamefont {Hussain},\ and\ \citenamefont {Jamil}}]{Majeed:2014kka}%
  \BibitemOpen
  \bibfield  {author} {\bibinfo {author} {\bibfnamefont {B.}~\bibnamefont
  {Majeed}}, \bibinfo {author} {\bibfnamefont {S.}~\bibnamefont {Hussain}},\
  and\ \bibinfo {author} {\bibfnamefont {M.}~\bibnamefont {Jamil}},\ }\href
  {https://doi.org/10.1155/2015/671259} {\bibfield  {journal} {\bibinfo
  {journal} {Adv. High Energy Phys.}\ }\textbf {\bibinfo {volume} {2015}},\
  \bibinfo {pages} {671259} (\bibinfo {year} {2015})},\ \Eprint
  {https://arxiv.org/abs/1411.4811} {arXiv:1411.4811 [gr-qc]} \BibitemShut
  {NoStop}%
\bibitem [{\citenamefont {Pugliese}\ \emph {et~al.}(2011)\citenamefont
  {Pugliese}, \citenamefont {Quevedo},\ and\ \citenamefont
  {Ruffini}}]{article}%
  \BibitemOpen
  \bibfield  {author} {\bibinfo {author} {\bibfnamefont {D.}~\bibnamefont
  {Pugliese}}, \bibinfo {author} {\bibfnamefont {H.}~\bibnamefont {Quevedo}},\
  and\ \bibinfo {author} {\bibfnamefont {R.}~\bibnamefont {Ruffini}},\ }\href
  {https://doi.org/10.1103/PhysRevD.83.104052} {\bibfield  {journal} {\bibinfo
  {journal} {Phys. Rev. D}\ }\textbf {\bibinfo {volume} {83}} (\bibinfo {year}
  {2011})}\BibitemShut {NoStop}%
\bibitem [{\citenamefont
  {Grabiner}(1999)}]{doi:10.1080/00029890.1999.12005131}%
  \BibitemOpen
  \bibfield  {author} {\bibinfo {author} {\bibfnamefont {D.~J.}\ \bibnamefont
  {Grabiner}},\ }\href {https://doi.org/10.1080/00029890.1999.12005131}
  {\bibfield  {journal} {\bibinfo  {journal} {The American Mathematical
  Monthly}\ }\textbf {\bibinfo {volume} {106}},\ \bibinfo {pages} {854}
  (\bibinfo {year} {1999})},\ \Eprint
  {https://arxiv.org/abs/https://doi.org/10.1080/00029890.1999.12005131}
  {https://doi.org/10.1080/00029890.1999.12005131} \BibitemShut {NoStop}%
\bibitem [{\citenamefont {Anderson}\ \emph {et~al.}(1998)\citenamefont
  {Anderson}, \citenamefont {Jackson},\ and\ \citenamefont
  {Sitharam}}]{ruleOfSings}%
  \BibitemOpen
  \bibfield  {author} {\bibinfo {author} {\bibfnamefont {B.}~\bibnamefont
  {Anderson}}, \bibinfo {author} {\bibfnamefont {J.}~\bibnamefont {Jackson}},\
  and\ \bibinfo {author} {\bibfnamefont {M.}~\bibnamefont {Sitharam}},\
  }\href@noop {} {\bibfield  {journal} {\bibinfo  {journal} {The American
  Mathematical Monthly}\ }\textbf {\bibinfo {volume} {105}},\ \bibinfo {pages}
  {447} (\bibinfo {year} {1998})}\BibitemShut {NoStop}%
\bibitem [{\citenamefont {Tsukamoto}(2021)}]{PhotonSpherePhysRevD.104.124016}%
  \BibitemOpen
  \bibfield  {author} {\bibinfo {author} {\bibfnamefont {N.}~\bibnamefont
  {Tsukamoto}},\ }\href {https://doi.org/10.1103/PhysRevD.104.124016}
  {\bibfield  {journal} {\bibinfo  {journal} {Phys. Rev. D}\ }\textbf {\bibinfo
  {volume} {104}},\ \bibinfo {pages} {124016} (\bibinfo {year}
  {2021})}\BibitemShut {NoStop}%
\bibitem [{\citenamefont {Znajek}(1976)}]{ZNAJEK1976}%
  \BibitemOpen
  \bibfield  {author} {\bibinfo {author} {\bibfnamefont {R.}~\bibnamefont
  {Znajek}},\ }\href {https://doi.org/10.1038/262270a0} {\bibfield  {journal}
  {\bibinfo  {journal} {Nature}\ }\textbf {\bibinfo {volume} {262}},\ \bibinfo
  {pages} {270} (\bibinfo {year} {1976})}\BibitemShut {NoStop}%
\bibitem [{\citenamefont {Frolov}(2012)}]{Frolov:2011ea}%
  \BibitemOpen
  \bibfield  {author} {\bibinfo {author} {\bibfnamefont {V.~P.}\ \bibnamefont
  {Frolov}},\ }\href {https://doi.org/10.1103/PhysRevD.85.024020} {\bibfield
  {journal} {\bibinfo  {journal} {Phys. Rev. D}\ }\textbf {\bibinfo {volume}
  {85}},\ \bibinfo {pages} {024020} (\bibinfo {year} {2012})},\ \Eprint
  {https://arxiv.org/abs/1110.6274} {arXiv:1110.6274 [gr-qc]} \BibitemShut
  {NoStop}%
\bibitem [{\citenamefont {H.~Se\c{c}uk}(2020)}]{HalukSecuk:2020oos}%
  \BibitemOpen
  \bibfield  {author} {\bibinfo {author} {\bibfnamefont {M.}~\bibnamefont
  {H.~Se\c{c}uk}},\ }\href {https://doi.org/10.1103/PhysRevD.102.068501}
  {\bibfield  {journal} {\bibinfo  {journal} {Phys. Rev. D}\ }\textbf {\bibinfo
  {volume} {102}},\ \bibinfo {pages} {068501} (\bibinfo {year}
  {2020})}\BibitemShut {NoStop}%
\end{thebibliography}%
\bibliographystyle{apsrev4-2}

\end{document}